\documentclass[prb,twocolumn,notitlepage,superscriptaddress]{revtex4-1}

\usepackage[left=1.8cm,right=1.8cm,top=1.85cm,bottom=2cm]{geometry}

\usepackage{amsmath}
\usepackage{amssymb}
\usepackage{graphicx}
\usepackage{stackrel}
\usepackage{ulem}
\usepackage{subfigure}

\usepackage{multirow}
\usepackage{units}
\usepackage{float}

\renewcommand\refname{References and Notes}

\newcounter{lastnote}

\renewcommand{\k}{\mathbf{k}}
\newcommand{\kp}{\mathbf{k}.\mathbf{p}}

\newcommand{\K}{{\mathbf{K}}}
\newcommand{\KP}{{\mathbf{K}'}}
\newcommand{\Ch}{\mathcal{C}}
\newcommand{\ketbra}[2]{\left|#1\middle>\middle<#2\right|}

\newcommand{\braOPket}[3]{\left<#1\middle|#2\middle|#3\right>}

\newcommand{\ket}[1]{\left|#1\right>}
\newcommand{\Ham}{\hat{H}}
\newcommand{\E}{\epsilon}
\newcommand{\mom}{\hat{\mathbf{p}}}
\newcommand{\SOCmom}{\hat{\boldsymbol{\pi}}}
\newcommand{\hc}{\textrm{h.c.}}

\newcommand{\IRseven}{A'}
\newcommand{\IRten}{E'}
\newcommand{\IReleven}{\bar{E}'}

\newcommand{\eeA}{\varOmega}
\newcommand{\eeC}{\bar{\varOmega}}

\usepackage{color}

\newcommand{\vp}{v_p}
\newcommand{\vd}{v_d}

\newcommand{\emphasize}[1]{#1}
\newcommand{\quoted}[1]{#1}

\renewcommand{\emph}[1]{\textit{#1}}
\renewcommand{\em}{\it}

\begin{document} 
\title{All-Optical Materials Design of Chiral Edge Modes in Transition-Metal Dichalcogenides} 
\author{Martin Claassen}
  \email[Author to whom correspondence should be addressed to: M. C. (\href{mailto:mclaassen@stanford.edu}{mclaassen@stanford.edu})
]{}
 \affiliation{Department of Applied Physics, Stanford University, CA 94305, USA} 
 \affiliation{Stanford Institute for Materials and Energy Sciences, SLAC \& Stanford University, CA 94025, USA}
\author{Chunjing Jia}
 \affiliation{Stanford Institute for Materials and Energy Sciences, SLAC \& Stanford University, CA 94025, USA} 
\author{Brian Moritz}
 \affiliation{Stanford Institute for Materials and Energy Sciences, SLAC \& Stanford University, CA 94025, USA}
 \affiliation{Department of Physics and Astrophysics, University of North Dakota,
Grand Forks, North Dakota 58202, USA}
\author{Thomas P. Devereaux}
 \affiliation{Stanford Institute for Materials and Energy Sciences, SLAC \& Stanford University, CA 94025, USA}
 \affiliation{Geballe Laboratory for Advanced Materials, Stanford University, Stanford, California 94305, USA}

\date{Aug. 15, 2016}

\begin{abstract}
Monolayer transition-metal dichalcogenides are novel materials which at low energies constitute a condensed-matter realization of massive relativistic fermions in two dimensions. Here, we show that this picture breaks for optical pumping - instead, the added complexity of a realistic materials description leads to a new mechanism to optically induce topologically-protected chiral edge modes, facilitating optically-switchable conduction channels that are insensitive to disorder. In contrast to graphene and previously-discussed toy models, the underlying mechanism relies on the intrinsic three-band nature of transition-metal dichalcogenide monolayers near the band edges. Photo-induced band inversions scale linearly in applied pump field and exhibit transitions from one to two chiral edge modes upon sweeping from red to blue detuning.  We develop an \textit{ab initio} strategy to understand non-equilibrium Floquet-Bloch bands and topological transitions, and illustrate for WS$_2$ that control of chiral edge modes can be dictated solely from symmetry principles and is not qualitatively sensitive to microscopic materials details.
\end{abstract}

\maketitle 

Manipulating materials properties far from equilibrium recently garnered significant attention, with experimental emphasis on transient melting, enhancement, or induction of electronic order \cite{Schmitt19092008,Fausti14012011,kim2012ultrafast,mankowsky2014nonlinear}. A more tantalizing aspect of the matter-light interaction regards the possibility to access dynamical steady states with distinct non-equilibrium phase transitions to affect electronic transport \cite{lindner2011floquet,wang2013observation,mahmood2016floquetvolkov,sie2014valley,kim2014ultrafast}. 
Conceptually simple, irradiation with a sufficiently broad pump pulse dresses the original electronic bands by multiples of the photon frequency, with electric dipole coupling resulting in an effective steady-state band structure; Floquet-Bloch theory then corresponds precisely to the classical limit of strong pump fields that are indistinguishable before and after photon absorption or emission. One paradigmatic model of such Floquet-Bloch bands is graphene \cite{oka2009photovoltaic,kitagawa2011transport,sentef2015}, where circularly-polarized light can break time-reversal symmetry to dynamically lift the Dirac point degeneracies. While Floquet-Bloch states were indeed observed recently via micro-wave pumping of Dirac cones on the surface of  topological insulators \cite{wang2013observation, mahmood2016floquetvolkov}, an extension to proper topological phase transitions is still well beyond experimental reach due to the tremendous required electric field: $\mathcal{E}_0$ to open a sizeable gap $\propto \mathcal{E}_0^2 / \Omega^3$ for above-bandwidth pump frequencies \cite{oka2009photovoltaic}. Conversely, experimentally realizable gaps at the Dirac points at lower pump frequencies \cite{kitagawa2011transport,sentef2015} or in semiconductor quantum wells \cite{lindner2011floquet} come at the price of resonant absorption, heating the sample, or worse, at the required pump strengths.

Viewed na\"ively as semiconducting analogs of graphene, trigonal-prismatic monolayers of MoS$_2$, MoSe$_2$, WS$_2$ and WSe$_2$ possess sizeable intrinsic band gaps due to broken inversion symmetry \cite{xiao2012coupled} that can be expected to sustain intense sub-gap pump pulses while limiting absorption for sufficient detuning from the band edge. Prior theoretical studies established that the band edges at $\K$ and $\KP$ are dominated by transition-metal $d$-orbitals, which split into three groups with irreducible representations (IRs) $A'\{ d_{3z^2-r^2} \}$, $E',\bar{E}'\{ d_{x^2-y^2} \pm i d_{2xy} \}$ and $E'',\bar{E}''\{ d_{xz} \pm i d_{yz} \}$ of the $C_{\rm 3h}$ point group  \cite{bromley1972band,mattheis1973band}. Generalizing graphene, these valleys are well-captured in equilibrium by a degenerate Kramers' pair of massive Dirac fermions, giving rise to valley-Hall \cite{xiao2012coupled} and spin-Hall \cite{feng2012intrinsic,shan2013spin} effects.

Out of equilibrium, dynamical breaking of time-reversal symmetry was demonstrated to lift the valley degeneracy for WS$_2$ and WSe$_2$ via off-resonant optical pumping with circularly-polarized light \cite{sie2014valley, kim2014ultrafast}. In this case, the selection rules for a massive Dirac fermion entail that the handedness of pump polarization selectively addresses either the $\K$ or $\KP$ valley, imparting an AC Stark shift on only one of the valleys. Analogously, the photo-excitation can selectively populate valleys, enabling spin and valley currents using circular or linear polarization \cite{yu2014nonlinear,shan2015optical,muniz2015all}.

\begin{figure*}[t]
\centering
\includegraphics[width=16cm]{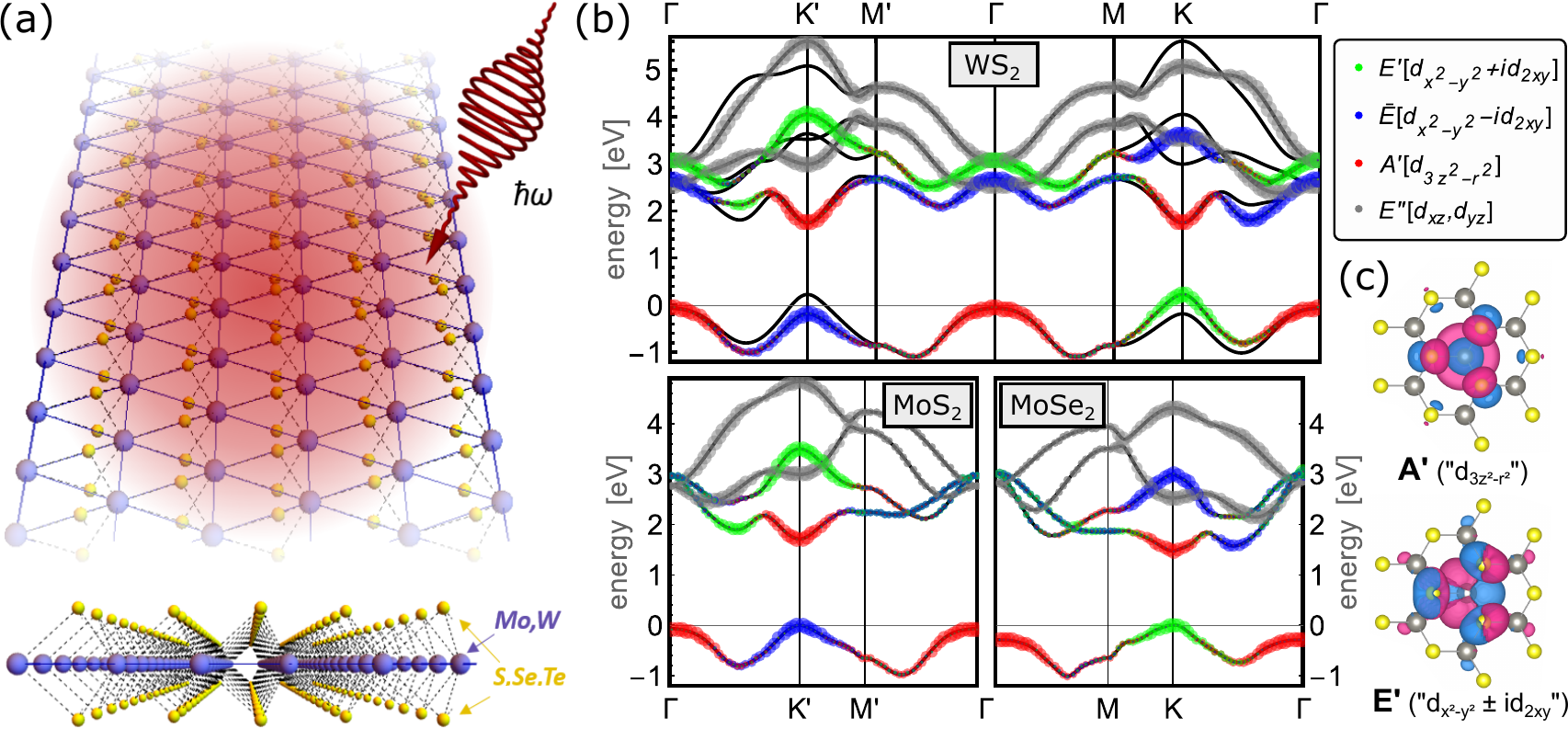}
\caption{\textbf{\textit{ab initio} electronic structure of transition-metal dichalcogenide monolayers.} (a) Setup: a circularly-polarized pump beam irradiates a monolayer transition-metal dichalcogenide ribbon. (b) Band structure and orbital content for MoS$_2$ and MoSe$_2$, as well as WS$_2$ with spin-orbit coupling, determined from \textit{ab initio} DFT calculations and downfolding onto localized Wannier orbitals. K and K' are related by time-reversal symmetry. (c) Isosurfaces highlight that the effective Wannier orbitals for WS$_2$, while localized on W, take into account orbital content extending to the S atoms as well as neighboring W atoms.} \label{fig:intro}
\end{figure*}

Even more tantalizingly, it was predicted that effective TMDC toy models -- graphene with a gap -- admit, in theory, an optically-induced quantum Hall effect with a single chiral mode localized at the sample edge. A high-frequency pump $\Omega\to\infty$ well above the bandwidth can in principle close and invert the equilibrium band gap at a single valley \cite{tahir2014photoinducedhall}; however, this requires a tremendous pump intensity.
Alternatively, it was proposed that a resonant pump beam can hybridize the massive Dirac fermion valence and conduction bands and thereby generate a single chiral edge mode \cite{sie2014valley} at lower pump strength. 

In this paper, we instead show that such a simple description fails to hold for optical pumping; here, the added complexity of a more realistic model of TMDC monolayers opens up a novel avenue to engineer a Floquet topological insulator in a realistic experimental setting. We argue that correctly addressing optical excitations necessitates a minimally three-band description near the band edges that leads to a frequeny-tunable mechanism to photo-induce one or two chiral edge modes. To understand the nature of concurrent Floquet band inversions at both $\K$ and $\KP$, we develop an \textit{ab initio} Floquet $\kp$ formalism that directly connects equilibrium density-functional theory (DFT) calculations with non-equilibrium Floquet theory. We illustrate these predictions for the example of a WS$_2$ ribbon, and present non-equilibrium ribbon spectra as well as an \textit{ab initio} characterization of the photo-induced valley topological band inversions. We find that control of chiral edge modes is determined solely by crystal symmetry and is insensitive to materials microscopics such as multi-photon processes or local inter-orbital dipole transitions that cannot be captured straightforwardly in a tight-binding model.

\begin{figure*}[t]
\centering
\includegraphics[width=16cm]{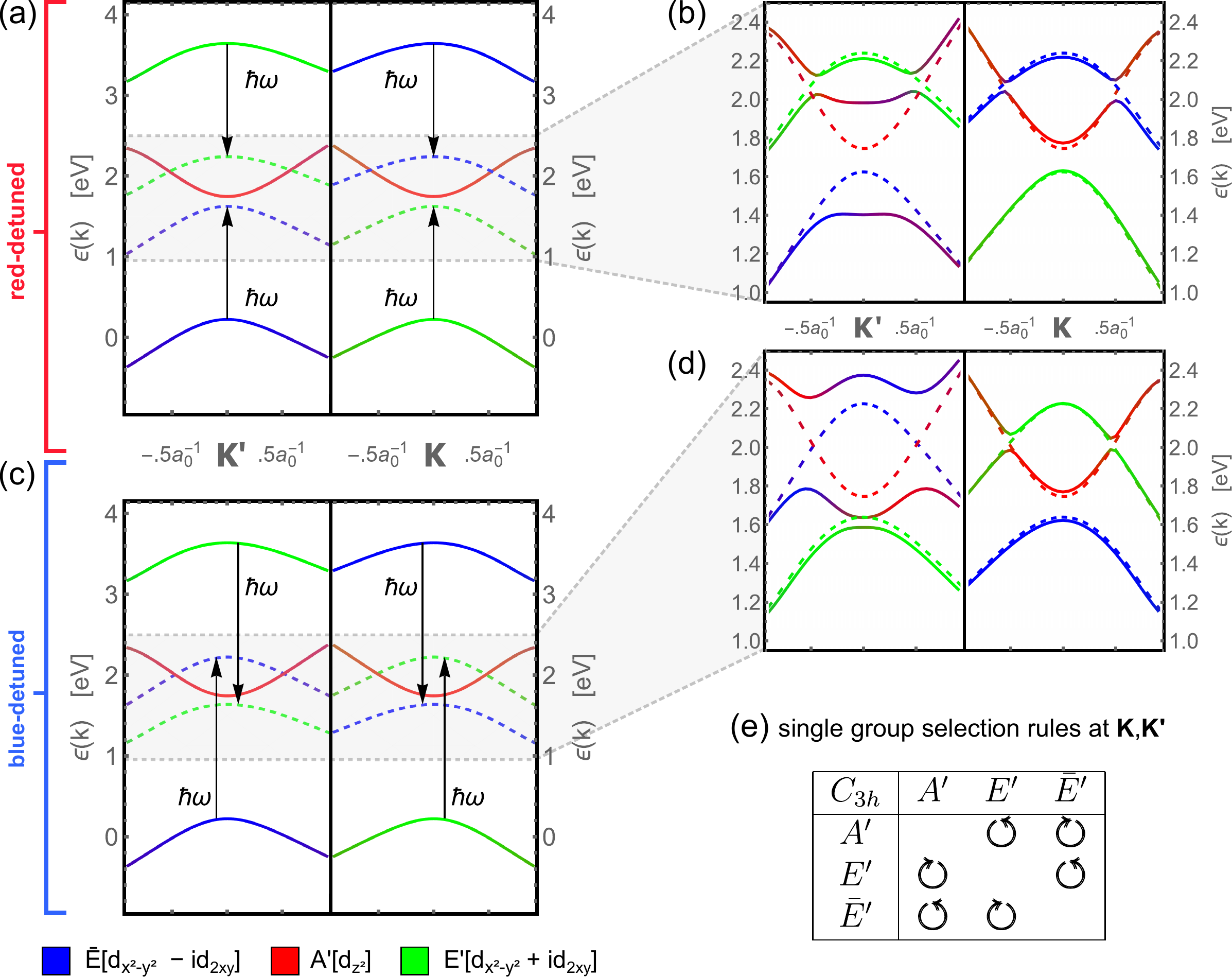}
\caption{\textbf{Photo-induced hybridization gaps and band inversion at $\K,\KP$.} Photon-dressed $\textit{ab initio}$ Floquet bands of Kramer's pair $\K$, $\KP$, for red (a,b) and blue (c,d) detuning. 
A red-detuned (blue-detuned) pump generically brings into resonance a ring of states of the $E'$ higher-energy band ($\bar{E}'$ valence band) with the $A'$ conduction band, while leaving the valence band (higher-energy band) off-resonant. Electric dipole coupling mixes the equilibrium orbital characters at $\K,\KP$ while lifting degeneracies between the conduction band and photon-dressed copies of the other bands (b,d). The ensuing photo-induced hybridization gap leads to topologically non-trivial band inversion at a single valley. (e) The corresponding $C_{\rm 3h}$ single group selection rules at $\K,\KP$ for circular polarization [see supplementary note 1, table 2 for a double group generalization]. }  \label{fig:hybridization}
\end{figure*}

\section*{Results}

\subsection*{Three-Band Nature of the Light-Matter Interaction} 
Central to this paper, dipole transitions to higher-lying bands, as determined directly from \textit{ab initio} calculations, and underlying symmetry considerations are crucial for a determination of photo-induced topological band inversions in TMDCs. To understand the breakdown of a description as \quoted{graphene with a gap}, consider a monolayer ribbon uniformly irradiated by circularly-polarized light, with collinear sample and polarization planes [Fig. \ref{fig:intro}(a)]. In graphene, the low-energy bands near the Fermi level are separated from higher-energy conduction bands by more than 10eV at $\K, \KP$~ \cite{kogan2014energy}, hence optical frequencies can be treated safely within the canonical low-energy Dirac model of $\boldsymbol{\pi}$ orbitals. In contrast, the band structures of prototypical TMDC monolayers possess a $E'$ band only $\sim 2$eV above the conduction band (CB) [Fig. \ref{fig:intro}(b)], and $E''$ bands in the same vicinity \cite{zahid2013generic,cappelluti2013tight,rostami2013effective,liu2013three}. Circularly-polarized light at close-to-bandgap pump frequencies therefore couples the $A'$ CB to \emphasize{both} the $E'$ valence band (VB) and the $\bar{E}'$ higher-energy conduction band (XB), while leaving the $E''$ bands decoupled in the absence of multipole transitions.

Consider first the case of slightly red-detuned pumping below the band edge [Fig. \ref{fig:hybridization}(a)]. Here, a ring of states from the higher-energy XB is brought into resonance with the CB, while simultaneously avoiding resonant coupling between the VB and CB. Central to experimental feasability, this regime can be expected to substantially limit absorption and heating.
At the band edge, $C_{\rm 3h}$ dipole selection rules [Fig. \ref{fig:hybridization}(e)] dictate that absorption of a photon couples transitions $A' \rightarrow \bar{E}' \rightarrow E' \rightarrow A'$. At $\KP$, the bare CB $\ket{m=0; A',{\rm CB}}$ (Floquet index $m$) couples to the XB dressed by a single emitted photon $\ket{m=-1; E',{\rm XB}}$ as well as the VB dressed by a single absorbed photon $\ket{m=+1; \bar{E}',{\rm VB}}$. Both transitions, though off-resonant, are energetically favorable, leading to a significant Stark shift at $\KP$ [Fig. \ref{fig:hybridization}(b)]. Conversely, at $\K$ the IRs of VB and XB are reversed. Here, the conduction band couples to the VB dressed by a single emitted photon $\ket{m=-1; E',{\rm VB}}$ as well as the XB dressed by a single absorbed photon $\ket{m=+1; \bar{E}',{\rm XB}}$. Both transitions are energetically unfavorable, leading to a negligible shift of the band edge. Slightly away from $\K$ and $\KP$, electric dipole coupling lifts the ring of degeneracy between the CB and XB and opens a photo-induced hybridization gap at both valleys [Fig. \ref{fig:hybridization}(b)], which scales \emphasize{linearly} with weak pump fields. Crucially, the resulting Floquet-Bloch bands exhibit a topological `band inversion' with the orbital character flipped close to the valley minimum, at \emphasize{both valleys} [Fig. \ref{fig:hybridization}(b)].

\begin{figure*}[!t]
\centering
\includegraphics[width=\textwidth]{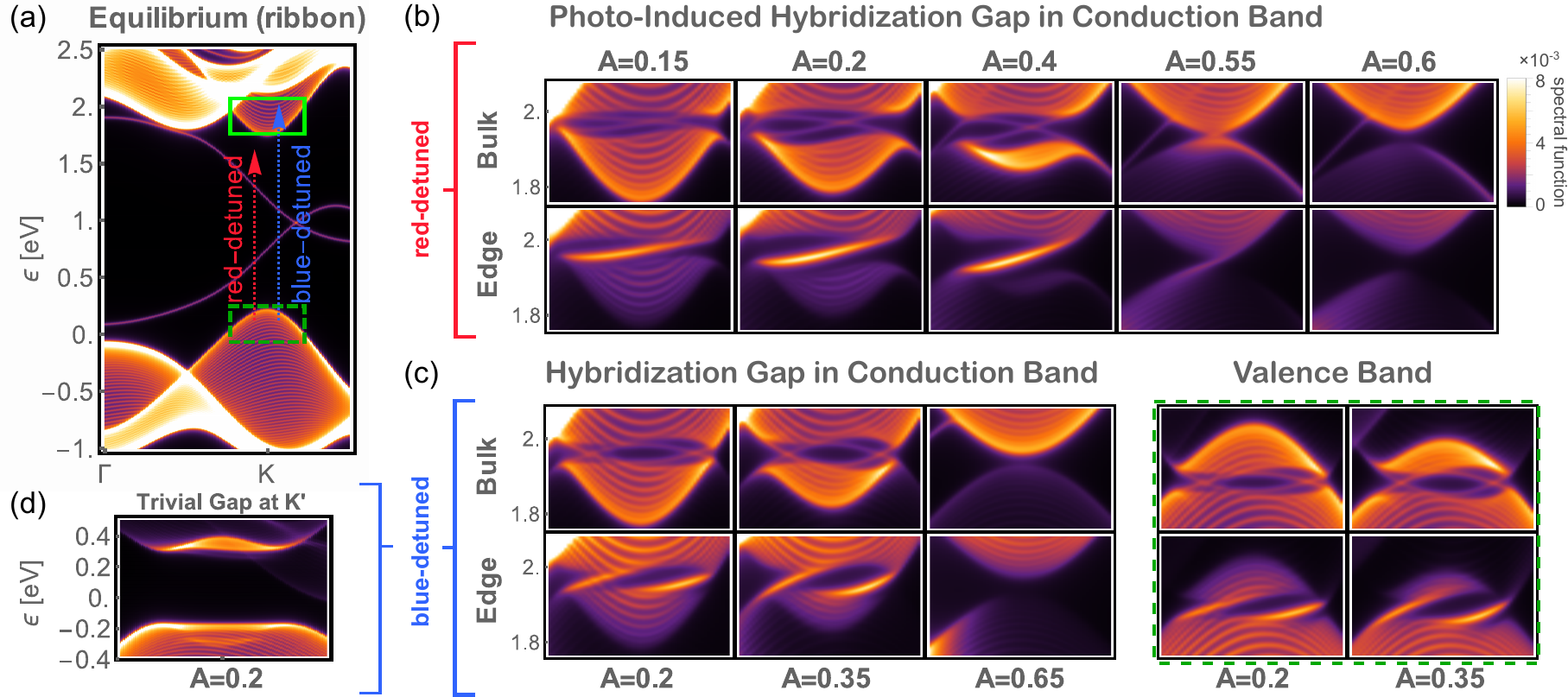}
\caption{\textbf{Single-particle spectra for a WS$_2$ ribbon irradiated with circularly-polarized light.} (a) In equilibrium, the ribbon hosts two trivial edge states localized at opposing edges. (b) A red-detuned pump field induces a hybridization gap at the bottom of the conduction band, which scales linearly in $A$ until closing again at a critical pump strength. (c) A blue detuned pump couples the equilibrium valence and conduction bands, inducing gaps both at the bottom of the conduction and top of the valence band, each hosting \emphasize{two} chiral edge modes. (d) Hybridization gaps at the opposite valley $\KP$ lead merely to a trivial band inversion, depicted here for the top of the valence band for blue detuning. }  \label{fig:ribbon}
\end{figure*}
\subsection*{Topology of Photo-Induced Band Inversions}

To discern whether the band inversions can be non-trivial, we devise effective Floquet two-band models of the hybridization gaps. We start from the generic description $\Ham_0 = \mom^2 /2m_0 + V(\mathbf{r})$ of a semiconductor in the absence of spin-orbit coupling and excitonic effects, where $V(\mathbf{r})$ is the crystal potential. 
In equilibrium, starting from an \textit{ab initio} Bloch eigenbasis at a single high-symmetry point, the dispersion and orbital content follow from canonical $\mathbf{k}.\mathbf{p}$ theory by perturbing in momentum deviation $\k$ under replacement $\mom \to \mom + \hbar \k$ \cite{winkler2003spin}. In the presence of a time-periodic field $\mathbf{A}(t) = A_0 \left[ \cos(\Omega t),~\sin(\omega t) \right]^\top$, a straight-forward generalization uses a \emphasize{Floquet eigenbasis} of the generic non-equilibrium problem $\Ham_0(t) =\left[\mom + e\mathbf{A}(t)\right]^2 /2m_0  + V(\mathbf{r})$. This basis can be obtained from density-functional theory calculations via knowledge of the equilibrium band energies and dipole transition matrix elements. Note that this Floquet $\mathbf{k}.\mathbf{p}$ theory is non-perturbative in the applied pump field and naturally accounts for multi-photon coupling to higher-energy CBs, XBs and deeper VBs, as well as local inter-orbital dipole transitions. In the following, $A = a_0 e \mathcal{E}_0 / (\hbar\Omega)$ denotes the dimensionless field strength with lattice constant $a_0 = 3.2~\textrm{\AA}$ and electric field $\mathcal{E}_0$; $A=0.1$ corresponds to $\mathcal{E}_0 \sim 47 ~\textrm{mV} \textrm{\AA}^{-1}$ for optical pump fields with $\Omega=1.5 ~\textrm{eV}$. An effective low-energy description of the photo-induced gaps can now be devised in Floquet basis in analogy to the equilibrium problem, by considering a perturbation in crystal momentum $\mom \to \mom + \hbar \k$ and downfolding onto effective two-band Floquet models using canonical L\"owdin perturbation theory.

Central to the robustness of this proposal, the form of these effective models is determined solely from symmetry and is universal to trigonal-prismatic TMDC monolayers. To see this, first consider $\K$. Here, the Floquet eigenbasis [Fig. \ref{fig:hybridization}(b)] $\ket{\Psi_1}$ ($\ket{\Psi_2}$) admixes $\ket{m=0;A',{\rm CB}}$ with $\ket{m=-1;E',{\rm VB}}$, $\ket{m=+1;\bar{E}',{\rm XB}}$ ($\ket{m=-1;\bar{E}',{\rm XB}}$ with $\ket{m=-2;A',{\rm CB}}$, $\ket{m=0;E',{\rm VB}}$), linear in field $A_0$.
Constrained by crystal symmetry, we find that the effective Floquet physics at $\K$ is generically determined by a $p$-$d$ Dirac model [see Methods]:
\begin{align}
	\Ham_\K(\k) &= \E_0(\k) + \left[\begin{array}{cc} \frac{1}{2}M - B|\k|^2 & \vp k_- - \vd k_+^2 \\ \vp k_+ - \vd k_-^2 & -\frac{1}{2}M + B |\k|^2 \end{array}\right]     \label{eq:HamEffK}
\end{align}
An additional purely dispersive term $\E_0(\k) = \Delta_0 + \Delta_2 |\k|^2$ breaks particle-hole symmetry. Crucially, the off-diagonal couplings $\vp, \vd$ are linear functions in field strength, suggesting a sizable photo-induced gap already at weak fields. While the parameters depend on the details of the Bloch states near the Dirac points, overall topological considerations can be gleaned simply from Eq. \ref{eq:HamEffK}. In the absence of $v_d$, Eq. (\ref{eq:HamEffK}) describes a conventional massive Dirac fermion, with $M$ ($B$) the Dirac (inverse band) mass, and $v_p$ the Dirac velocity. The orbital character exhibits a $p$-wave winding around $\K$ and mirrors the quantum anomalous Hall effect in Hg$_y$Mn$_{1-y}$Te quantum wells \cite{liu2008quantum}. Switching on $v_d$ imparts a trigonal distortion by reducing the continuous rotational symmetry around $\K$ to $C_3$, and introduces instead a `$d$-wave' winding in the limit $\sqrt{B/M} ~v_d \gg v_p$.

At $\KP$ interchanged IRs $E',\bar{E}'$ entail a strongly-admixed Floquet eigenbasis as well as a significant Stark shift. However selection rules forbid a coupling between $\ket{\Psi_1}, \ket{\Psi_2}$ linear in $\k$ --- instead, one finds that the two bands couple to quadratic order in $\sim A k_+ k_-$, via intermediate states $\ket{m=0;E', XB}$ or $\ket{m=-1;A', CB}$. The effective Hamiltonian for $\KP$ in this case generically reads
\begin{align}
	\Ham_\KP(\k) &= \E_0'(\k) + \left[\begin{array}{cc} \frac{1}{2}M' - B' |\k|^2 & v' |\k|^2 \\ v' |\k|^2 & -\frac{1}{2}M' + B' |\k|^2 \end{array}\right]       \label{eq:HamEffKP}
\end{align}
with $v'$ a rotationally-symmetric band mixing term.

At $\K$ ($\KP$), the band ordering is inverted when $M/B > 0$ ($M'/B' > 0$). If the orbital character of Floquet-Bloch bands in the remaining Brillouin zone is sufficiently benign, we can draw conclusions on the global topology by understanding separately the band inversions at $\K$ and $\KP$. Rewriting Eqs. (\ref{eq:HamEffK}), (\ref{eq:HamEffKP}) in terms of Pauli matrices $\Ham = \E_0(\k) + \boldsymbol{\sigma} \cdot \mathbf{d}(\k)$, the local Berry curvature follows from the winding $\mathcal{F}(\k) = \frac{1}{2} \hat{\mathbf{d}}(\k) \cdot (\partial_{k_x} \hat{\mathbf{d}}(\k) \times \partial_{k_y} \hat{\mathbf{d}}(\k) )$ with $\hat{\mathbf{d}}(\k) = \mathbf{d}(\k)/|\mathbf{d}(\k)|$. One can see by inspection that the absence of $\hat{\boldsymbol{\sigma}}_y$ in Eq. (\ref{eq:HamEffKP}) enforces $\mathcal{F}(\k) = 0$ ; therefore, the photo-induced band inversion around $\KP$ is necessarily trivial. 
Conversely, the band inversion at $\K$ is topological and triggers a change in the Chern number $\Ch = \frac{1}{2\pi} \int_{\mathbb{R}} d\mathbf{k}~ \mathcal{F}(\k)$, which captures the contribution to topology arising from the band inversion in the vicinity of $\K$ [see Supplementary Note 3, Supplementary Figure 1].

Consider first the limit of a massive Dirac model with $v_d = 0$. In this case, the Chern number changes from $\Ch = 0$ for $B/M < 0$ to $\Ch = \pm 1$ for $B/M > 0$, inducing a single chiral edge mode, spanning the photo-induced hybridization gap, and localized at the boundary of a uniformly illuminated sample.
Switching on $v_d$ introduces a trigonal distortion of the Floquet-Bloch bands around $\K$ up to a critical strength $v_d^2 = B/M v_p^2$, at which the Floquet-Bloch bands close the gap at \emphasize{three points} away from $\K$, related by $C_3$. Correspondingly, this topological transition changes $\Ch$ by $3$, to $\Ch=\mp 2$, entailing not one but \emphasize{two chiral edge modes} at the sample boundary.

\begin{figure*}[t]
\centering
\includegraphics[width=\textwidth]{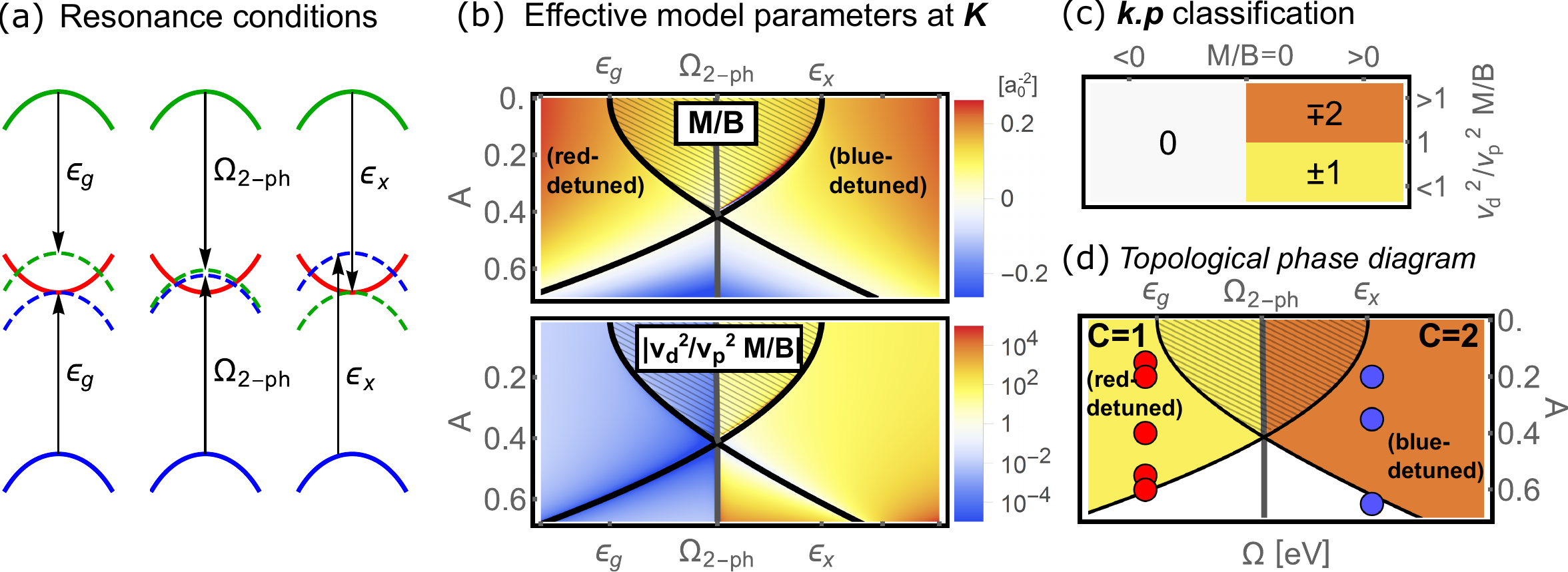}
\caption{\textbf{Topological Band Inversions and Floquet $\kp$ Theory in a Wannier Three-Orbital Model of WS$_2$.} (a) The relevant resonances when moving from red to blue detuning. (b) Effective $\kp$ parameterization of the band inversion at $\K$ [Eqn. \ref{eq:HamEffK}] starting from the three-orbital Wannier tight-binding model. (c) sketches the corresponding topological classification. At low pump strengths, the sign of $M/B$ $a_0^{2}$ is negative in the non-trivial inverted regime. The system transitions from one (red-detuned) to two (blue-detuned) chiral edge modes when $\left|\left(v_d/v_p\right)^2 \frac{M}{B}\right| > 1$. The effective $\mathbf{k}.\mathbf{p}$ model parameterized in (b) accounts for the Floquet state adiabatically derived from the CB, as well as the Floquet state deriving from the VB dressed by one absorbed photon, for red detuning, or from the higher-energy $E'$ XB dressed by one emitted photon, for blue detuning. This entails the sharp transition when sweeping the pump frequency across the two-photon resonance between VB and XB (a). Within the shaded areas, all three bands come into resonance. Thick black lines indicate degeneracies in the Floquet spectrum. (d) A corresponding numerical calculation of the global Chern number of the tight-binding model mirrors the $\kp$ analysis. Yellow (orange) denotes $\Ch=1$ ($\Ch=2$) as defined in (c), circles indicate parameters used for Fig. \ref{fig:ribbon}.}  \label{fig:kpWannier}
\end{figure*}

\subsection*{Red vs. Blue Detuning}

To get an understanding of the mechanism that might trigger this transition, note that while
the relevant Floquet basis is predominantly built from only the $m=0$ CB and $m=-1$ XB for a red-detuned pump, the $p$-wave coupling $v_p$ between the two is necessarily mediated via the VB (or other bands of equivalent IR),
highlighting the necessity of a minimal three-band description [see Supplementary Note 2]. We stress that such an effective Dirac-like contribution is generically impossible to obtain starting from a two-band equilibrium description as a massive Dirac fermion. In contrast, the $d$-wave term results from direct coupling between CB and XB. 
Strong optical absorption in TMDC monolayers indicates a large dipole transition matrix element 
between VB and CB, suggesting that a sufficiently red-detuned pump will generically reach the $C=1$ phase. Note that the asymmetry between $\K,\KP$ stems from the choice of polarization, and reverses for opposite handedness of the circularly polarized pump beam.

In contrast, consider the opposite regime of a sufficiently blue-detuned pump [Fig. \ref{fig:hybridization}(c)]. Here, a ring of VB states is brought into resonance with the CB near $\K,\KP$ while pushing the photon-dressed XB into the equilibrium band gap. Electric dipole coupling again opens photo-induced hybridization gaps, both at the bottom of the CB and top of the VB [Fig. \ref{fig:hybridization}(d)], and the symmetry analysis mirrors the discussion of the red-detuned case above, leading to equivalent effective Hamiltonians at $\K,\KP$ (\ref{eq:HamEffK}), (\ref{eq:HamEffKP}). However, linear in $\k$ coupling between the $m=0$ CB and the $m=+1$ VB is now necessarily mediated via the XB (or other $\bar{E}'$ bands separated further in energy), whereas the $d$-wave term $v_d$ follows directly from dipole coupling between VB and CB, dominating over $v_p$. One can thus generically expect a frequency-tunable Floquet Chern insulator in monolayer TMDCs.

\subsection*{\textit{Ab Initio} Floquet Analysis of WS$_2$ Monolayers}

To illustrate these predictions, consider WS$_2$ as a prototypical TMDC monolayer. First, we perform \textit{ab initio} DFT calculations to derive an effective minimal tight-binding model of three $A',E',\bar{E}'$ Wannier orbitals localized on the W transition metal [Fig. \ref{fig:intro}(b,c)]. The resulting Floquet spectrum on a ribbon is depicted in Fig. 3, calculated from the period-averaged single-particle spectral function [see Methods]. In equilibrium [Fig. \ref{fig:ribbon}(a)], WS$_2$ already hosts a pair of trivial edge states in analogy to zigzag edges of graphene, with right (left) propagating modes at the $\K~(\KP)$ point that span the band gap. A weak, red-detuned pump field opens a hybridization gap at the bottom of the CB, spanned by a \emphasize{single} chiral mode at $\K$ [Fig. \ref{fig:ribbon}(b)], localized at the sample edge. The photo-induced gap scales linearly with weak $A$, but closes and reopens at a critical pump strength, transitioning again to a trivial phase without chiral modes. Conversely, for a blue-detuned pump [Fig. \ref{fig:ribbon}(c)] a \emphasize{second} chiral edge mode appears, spanning the hybridization gaps both at the bottom of the CB and the top of the VB.

The appearances of edge modes in ribbon spectra are in excellent agreement with effective model parameters [Eqs. (\ref{eq:HamEffK}), (\ref{eq:HamEffKP})] derived from the Wannier tight-binding description. Fig. \ref{fig:kpWannier}(b) depicts $M/B$ and the ratio of $p-$/$d-$wave couplings that determine the Chern number for the band inversion at $\K$ [Fig. \ref{fig:kpWannier}(c)], in perfect correspondence with a rigorous calculation of the $2+1$D Floquet winding number \cite{rudner2013anomalous} of the driven Wannier tight-binding model [Fig. \ref{fig:kpWannier}(d), see Supplementary Note 3], with the circularly-polarized pump entering via Peierls substitution. For weak fields, deep within both the red- and blue-detuned regimes, the sign of the Dirac $M$ and band $B$ mass are equal in the topologically-nontrivial phase. Here, $C\neq 0$, and the Chern number follows from trigonal distortion and changes from $C=1$ for red detuning to $C=2$ for blue detuning. Increasing $A$ instead closes and reopens the Floquet gap at $\K$, flips the sign of $M$ and uninverts the bands to reach a trivial phase with $C=0$. We note that this picture breaks down for intense pump fields with energy scales on the order of the equilibrium band gap; here, additional topological phase transitions can arise [Supplementary Figure 2, Note 4].

\begin{figure}[b]
\centering
\includegraphics[width=\columnwidth]{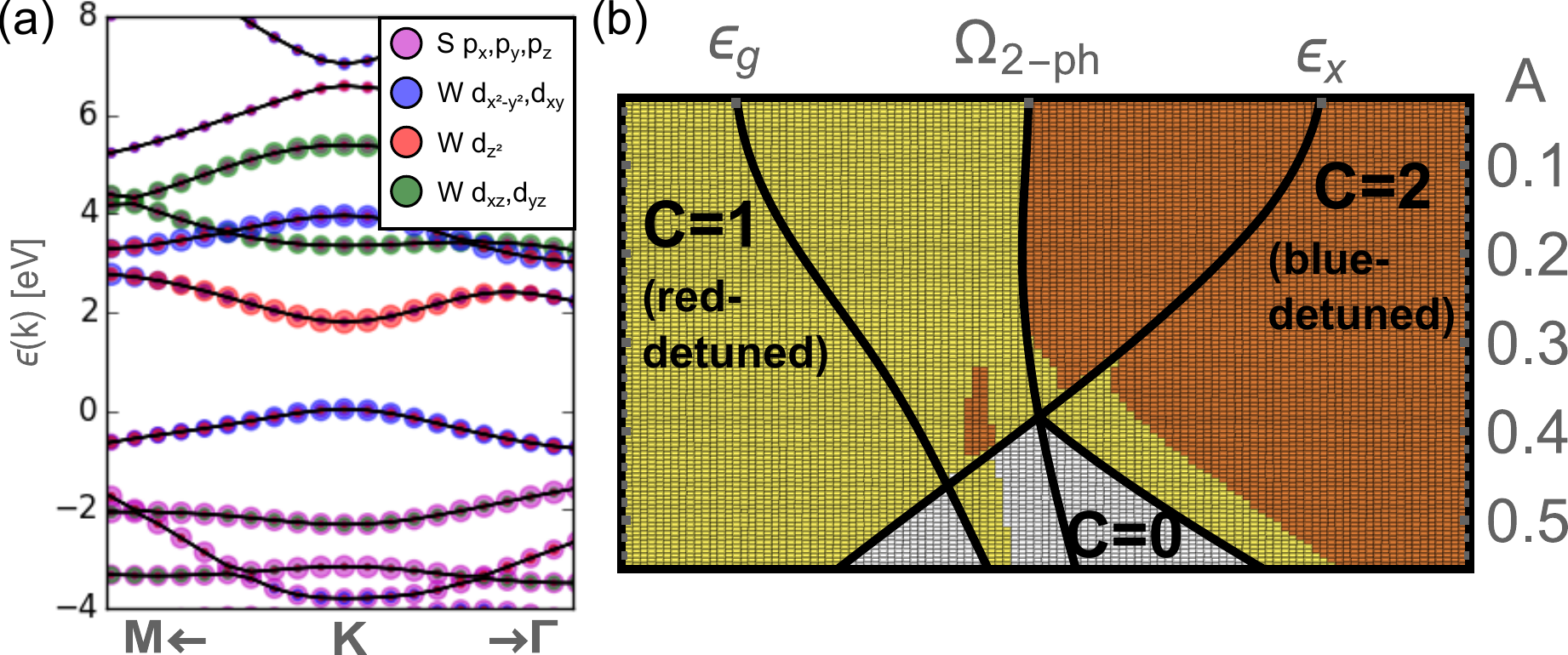}
\caption{\textbf{\textit{Ab initio} Floquet $\boldsymbol{k}.\boldsymbol{p}$ Classification of Photo-Induced Band Inversions in WS$_2$.} (a) Starting from a 185-band first-principles description of the $\K$ point, multi-photon processes involving deep core levels and higher-energy as well as local dipole transitions can be taken into account exactly. (b) Recomputing the effective Floquet $\kp$ two-band model of the photo-induced band inversion recovers a phase diagram that is qualitatively similar to the three-band tight-binding description [Fig. \ref{fig:kpWannier}(d)], suggesting that the photo-induction of one or two chiral edge modes for appropriate tuning of the pump laser is robust to multi-photon processes at weak pump fields. Strong pump-field deviations indicate a non-trivial admixing of higher-energy bands.} \label{fig:kpAbInitio}
\end{figure}

Having checked the validity of Floquet $\kp$ theory in the tight-binding model, we now turn to the full \textit{ab initio} problem.
To quantify the effects of multi-photon resonances, as well as local inter-orbital dipole transitions not captured in a tight-binding model, we consider an \textit{ab initio} 185-band description of band energies and dipole transition matrix elements at $\K$ and $\KP$ and calculate the model parameters of Eqns. (\ref{eq:HamEffK}), (\ref{eq:HamEffKP}), taking into account up to four-photon processes. The bands closest to the equilibrium gap are depicted in Fig. \ref{fig:kpAbInitio}(a). The resulting $\kp$ classification is depicted in Fig. \ref{fig:kpAbInitio}(b). Crucially, while the resonance lines distort due to effects not accounted for in the tight-binding model, the frequency-dependent switch from $C=1$ to $C=2$ as well as the reclosing of the hybridization gap and transition back to a trivial regime with increasing pump strength remains qualitatively similar. This suggests that the mechanism of photo-induced chiral edge modes described in this work is largely robust at weak fields to the microscopic details of the material.

\section*{Discussion}

A key challenge for a condensed-matter realization of Floquet topological insulators regards driving the system strongly to induce the required changes to the equilibrium band structure while simulteanously mitigating the inevitable absorption and heating in such a scheme. However, a common thread for pioneering works on graphene \cite{kitagawa2011transport,sentef2015}, semiconductor quantum wells \cite{lindner2011floquet} and topological insulator surface states \cite{wang2013observation,mahmood2016floquetvolkov} is that these require either high-frequency pumping at tremendous pump strengths or resonant pumping, thereby injecting substantial energy into the system and heating the sample. Recent work has tried to tackle these problems by discerning whether special couplings to bosonic or fermionic heat sinks \cite{dehghani2014dissipative,iadecola2015occupation,seetharam2015population} can dissipate enough energy and nevertheless stabilize a Floquet steady-state ensemble. Analogously, the blue detuned regime entails a ring of resonance between CB and VB, leading to enhanced absorption.
Conversely, and central to the experimental feasability of this proposal, red-detuned pumping of monolayer TMDCs circumvents these issues by entirely avoiding resonant coupling between VB and CB while nevertheless inducing non-trivial band topology by virtue of the minimally three-band nature of electron-photon coupling. Na\"ively, this can be understood by noting that, for weak recombination rates, the rate of carrier photo-excitation scales as $g^2 A^2 / \delta^2$ whereas the photo-induced hybridization gap scales linearly in pump strength $A$ [see Methods], where $\delta$ is the laser detuning from resonance and $g$ an effective dipole matrix element. 
Nevertheless, this residual photo-excited population will lead to broadening of the Floquet-Bloch bands due to recombination and electron-phonon scattering and set a lower bound on observable bulk hybridization gaps, while at the same time serving as a necessary ingredient to reach a steady-state population. Extending topological characterizations to open quantum systems in the limit of strong dissipation remains an interesting topic for future study \cite{budich2015topology}.

Smoking-gun evidence for the presence of chiral edge modes necessitates either measurement of the sample edge via nano angle-resolved photo emission (nano-ARPES) and local admittance probes such as microwave impedance spectroscopy, or direct transport measurements, scaling of conductance with sample length, where care must be taken regarding coupling between leads and the chiral Floquet edge mode \cite{kundu2013sumrule,farrell2015photon}. Conversely, ARPES stands as an immediate tool to verify the predicted photo-induced hybridization gaps in the 2D bulk, as a function of pump strength and when sweeping from red to blue detuning. {For red detuning, the CB can be enhanced in two-photon photoemission, in analogy to probing the unoccupied higher-energy topological surface states of Bi$_2$Se$_3$ \cite{sobota2011unoccupied}.} An interesting follow-up question concerns potential matrix element dependencies of photo-emission from the W $d$-orbitals, in order to directly observe and characterize the band inversions at $\K,\K'$ via bulk measurement. 

The guiding theme of this work has been to build a bridge between the rapidly developing field of monolayer transition-metal dichalcogenides and topological phase transitions out of equilibrium, to provide a route towards achieving the latter in an experimentally-attainable setting. We have shown that the three-band nature of the valleys in prototypical WS$_2$ leads to a new mechanism to `switch' on or off one or two chiral edges with near band-gap optical irradiation. The resulting photo-induced gap in the single-particle spectrum scales linearly with pump strength, suggesting substantial energy scales already at low fields, while simultaneously ensuring minimal heating with sufficient detuning from the band edge. Our theoretical analysis of the out-of-equilibrium valley band inversions connects directly with equilibrium \textit{ab initio} calculations, whereas the ensuing topology of Floquet-Bloch bands relies purely on generic symmetry arguments, suggesting that the predictions are robust to microscopic detail and should be observable in a range of monolayer TMDC materials. Finally, our first-principles and theoretical analysis provides a promising strategy to predict and design topological states out of equilibrium in other semiconductor materials.

\section*{Methods}

\subsection*{\textit{ab initio} Calculations}

\textit{ab initio} calculations were performed in the framework of the Perdew-Burke-Ernzerhof (PBE) type generalized gradient approximation (GGA) of density functional theory (DFT) using the full-potential linearized augmented plane wave method implemented in Wien2k  \cite{blahaWIEN2k}. We consider a single monolayer of WS$_2$ with a $30{\textrm{\AA}}$ vacuum space perpendicular to the layer along the z-direction. The in-plane lattice constant and the S position have been relaxed by optimization of the total energy and total force, respectively. For electronic structure calculations, we utilized a 15$\times$15$\times$1 $k$-space grid. Momentum matrix element calculations were performed using the OPTIC package implemented in Wien2k, with a 60$\times$60$\times$1 $k$-space grid. Maximally-localized Wannier functions (MLWFs) for the five W 5$d$ orbitals were obtained using wien2wannier \cite{kunes2010} and Wannier90 \cite{Mostofi2008} with initial projections set to the spherical harmonics $Y_{2m}$ ($m$ = -2, -1, 0, 1, 2). Due to the symmetry of the hexagonal lattice, the calculated Hamiltonian in the new Wannier basis naturally decouples into the two standard subspaces \{ $d_{x^{2}-y^{2}} \pm i\,d_{2xy}$, $d_{3z^{2}-r^{2}}$ \} and \{ $d_{xz} \pm i d_{yz}$ \}.

\subsection*{Floquet theory of the single-particle spectrum on a ribbon}

Floquet theory captures the effective steady states that arise from a time-dependent (quasi-)periodic modulation. Consider a Hamiltonian $\Ham(t) = \Ham(t+\frac{2\pi}{\Omega})$  with a periodic time dependence with frequency $\Omega$. Then, solutions of the time-dependent Schr\"odinger equation for $\Ham(t)$ can be written as $\Phi(t) = e^{-i\E t} \sum_m u_m e^{im\Omega t}$, where $\E$ is the Floquet quasi-energy, and $u_m$ are Fourier coefficients of the time-periodic part of the wave function. Substitution of $\Phi(t)$ into Schr\"odinger's equation recasts the time-dependent problem as an effective time-independent Floquet problem: the Floquet states can be found by finding eigenstates of the Floquet Hamiltonian
\begin{align}
	\Ham_F = \sum_{mm'} \left[ \Ham_{m-m'} + m\Omega \delta_{m,m'} \right] \ketbra{m}{m'},
\end{align}
where $\Ham_{m-m'} = \frac{\Omega}{2\pi} \int_0^{2\pi/\Omega} \Ham(t) e^{-i(m-m')\Omega t}$ are the Fourier expansion coefficients of $\Ham(t)$. If the original Hamiltonian has a static eigenbasis $\ket{\alpha}$, then the eigenstates of $\Ham_F$ can be written as $\ket{\lambda} = \sum_m u_{m\alpha}^{(\lambda)} \ket{\alpha}\otimes\ket{m}$, with the original time-dependent eigenstates of $\Ham(t)$ becoming $\ket{\lambda(t)} = e^{-i\E_\lambda t} \sum_m u_{m\alpha}^{(\lambda)} ~e^{im\Omega t} \ket{\alpha}$.
The next step is to connect back to observables of the original fermion operators. In the main text, we consider the spectral function
\begin{align}
	A(\omega,x) = -2\textrm{Im} &\left\{ \frac{\Omega}{2\pi} \sum_{\alpha\alpha'} \int_0^{2\pi/\Omega} dT \int_{-\infty}^{\infty} d\tau e^{i\omega\tau} ~ \times \right. \notag\\
	&~~~~~\times \left. \vphantom{\int_0^{2\pi/\Omega}} G_{\alpha\alpha'}^R\left(x,T+\frac{\tau}{2};x,T-\frac{\tau}{2}\right) \right\}
\end{align}
with the retarded Green's function defined as $G_{\alpha\alpha'}^R(x,t;x',t') = -i\theta(t-t') \left< \left\{ \Psi_\alpha(x,t), \Psi^\dag_{\alpha'}(x',t') \right\}\right>$. Rewriting the fermion operators $\Psi_\alpha(x,t)$ in Floquet basis, one finally arrives at the Floquet spectral function
\begin{align}
	A(\omega,x) \sim \sum_{m\lambda\alpha} \left| u_{mx\alpha}^{(\lambda)} \right|^2 \frac{\Gamma}{(\omega-\E_\lambda+m\Omega)^2 + \Gamma^2}
\end{align}
where $\Gamma$ is a phenomenological broadening of the spectrum.

\subsection*{Floquet $\mathbf{k}.\mathbf{p}$ Theory and Effective Hamiltonians at $\K,\KP$}

Consider a generic time-dependent starting point for $\K$ (and equivalently for $\KP$)
\begin{align}
	\Ham(t) &= \frac{1}{2m_0} [\mom + \hbar\k + e\mathbf{A}(t)]^2 + V(\mathbf{r}) \notag\\
	&+ \frac{\hbar}{4m_0^2c_0^2} [\mom + \hbar\k + e\mathbf{A}(t)] \cdot \hat{\boldsymbol{\sigma}} \times \boldsymbol{\nabla}V(\mathbf{r}) 
\end{align}
where $\k$ is the deviation in crystal momentum from $\K$ or $\KP$, with a respective shift to the $\K$ or $\KP$ point absorbed in $\mom$. We consider circularly-polarized light with $\mathbf{A}(t) = A \left[ \cos(\Omega t),~ \sin(\Omega t) \right]^\top$. Now decompose $\Ham(t)$ into equilibrium [$\Ham_{0,\textrm{eq}}$] and non-equilibrium [$\Ham_{0,\textrm{pump}}(t)$] constituents that determine the eigenbasis at $\K,\KP$, as well as a perturbation in $\k$ [$\Ham'(t)$]:
\begin{align}
	\Ham_{0,\textrm{eq}} &= \frac{\mom^2}{2m_0} + V(\mathbf{r}) + \frac{\hbar}{4m_0^2c_0^2} ~\mom \cdot \hat{\boldsymbol{\sigma}} \times \boldsymbol{\nabla}V(\mathbf{r}) \\
	\Ham_{0,\textrm{pump}}(t) &= \frac{e}{m_0} \mathbf{A}(t) \cdot \SOCmom + \frac{e^2 [\mathbf{A}(t)]^2}{2m_0} \\
	\Ham'(t) &= \frac{\hbar^2 \k^2}{2m_0} + \frac{\hbar}{m_0} \k \cdot \SOCmom + \frac{\hbar e}{m_0} \k \cdot \mathbf{A}(t)
\end{align}
where $\SOCmom = \mom + \frac{\hbar}{4m_0^2c_0^2} \hat{\boldsymbol{\sigma}} \times \boldsymbol{\nabla}V(\mathbf{r})$.
In equilibrium, the (time-independent) eigenbasis $\ket{\alpha,n}$ of $\Ham_{0,\textrm{eq}}$ can be determined from \textit{ab initio} calculations and transforms according to $C_{\rm 3h}$, with $n, \alpha$ indexing the n\textsuperscript{th} band with IR $\alpha$. In the absence of the pump field $\mathbf{A}(t) = 0$, conventional $\k.\mathbf{p}$ theory proceeds by considering the deviation in Bloch momentum as a perturbation, described by $\Ham'$.

In Floquet $\k.\mathbf{p}$ theory, one instead starts from the exact \emphasize{Floquet eigenbasis} of $\Ham_0(t) = \Ham_{0,\textrm{eq}} + \Ham_{0,\textrm{pump}}(t)$ at $\K$ and $\KP$. Consider a single spin manifold, and for simplicity denote bands by the $C_{\rm 3h}$ single group IRs $A',E',\bar{E}'$ [see Supplementary Table 2, Supplementary Note 1 for equivalent double group identifications and selection rules in the full SOC problem]. The selection rules [Fig. \ref{fig:hybridization}(e), Supplementary Table 1] then entail that $\Ham_{0,\textrm{eq}} + \Ham_{0,\textrm{pump}}(t)$ involve transitions
\begin{align}
	\scriptstyle \cdots \longleftrightarrow \ket{m-1;E',n} \stackrel{g_{nn'}^{E'A'}}{\longleftrightarrow} \ket{m;A',n'} \stackrel{g_{n'n''}^{A'\bar{E}'}}{\longleftrightarrow} \ket{m+1;\bar{E}',n''} \longleftrightarrow \cdots \label{eq:FloquetSchematics}
\end{align}
Here, $m$ ($n$) are Floquet (band) indices, and $g_{nn'}^{\alpha\alpha'} = \frac{\hbar}{2m_0}\braOPket{\alpha,n}{\left(\SOCmom_x - i \SOCmom_y\right)}{\alpha',n'}$ are the momentum matrix elements for allowed dipole transitions [Fig. \ref{fig:hybridization}(e)], obtained from \textit{ab initio} calculations. Employing a sufficiently large number of \textit{ab-initio}-determined Bloch states at $\K,\KP$, their dipole matrix elements and Floquet side bands, the Floquet eigenbasis at $\K,\KP$ can formally be determined exactly as functions of $A,\Omega$.

The effective two-band Hamiltonians [Eqns. \ref{eq:HamEffK},\ref{eq:HamEffKP}] described in the main text now follow via choosing two Floquet eigenstates for $\K$ and $\KP$ each, that are adiabatically connected to the $A=0$ CB with $m=0$ as well as the $m=-1$ XB ($m=+1$ VB) for red (blue) detuning, and employing L\"owdin perturbation theory to downfold $\Ham'(t)$ onto this two-state Floquet eigenbasis [a detailed derivation can be found in supplementary note 2].

Crucially, to distinguish $\K$ and $\KP$, note that their irreducible representations $E',\bar{E}'$ for VB, XB interchange. A simple way to arrive at the effective Hamiltonians (\ref{eq:HamEffK}), (\ref{eq:HamEffKP}) follows from observing for (\ref{eq:FloquetSchematics}) that the Floquet eigenbasis at $\K,\KP$ necessarily decomposes again into three IRs of the joint electron-photon problem, which are subsequently coupled by the Bloch momentum perturbation $\Ham'(t)$. In this picture, at $\K$, the IRs of the two Floquet basis states of (\ref{eq:HamEffK}) differ; hence off-diagonal coupling enters already at linear order $\sim k_{\pm}$. Note that this coupling can necessarily only arise in the present minimally three-band description [see supplementary material]. Conversely, at $\KP$ the IRs of the basis of (\ref{eq:HamEffKP}) are the same; off-diagonal coupling therefore necessarily enters only to quadratic order $\sim k_+ k_-$ and higher, leading to a trivial band inversion. 

The inclusion of full spin-orbit coupling does not qualitatively alter these conclusions. First, spin-flip terms weakly admix $E''$ bands of opposite spin \cite{Kormanyos2013,Gibertini2014}; however, the full crystal double group $\bar{C}_{\rm 3h}$ again decomposes into two spin-orbital manifold with equivalent selection rules and effective physics [supplementary note 1]. Second, the valley Zeeman shift simply leads to a shift of the relevant resonance energies. Similarly, while monolayer TMDCs have been shown to give rise to large excitonic binding energies \cite{Chernikov2014,Zhu2015}, in the context of our work their role is confined to shifting the relevant resonance energies, given appropriate tuning of the pump frequency.

\section*{Acknowledgements}

We gladly acknowledge helpful discussions with Alexander Kemper, Ruixing Zhang, Patrick Kirchmann, Tony Heinz, and Zhi-Xun Shen. This work was supported by the U. S. Department of Energy, Office of Basic Energy Science, Division of Materials Science and Engineering under Contract No. DE-AC02-76SF00515.  Computational resources were provided by the National Energy Research Scientific Computing Center supported by the Department of Energy, Office of Science, under Contract
No. DE- AC02-05CH11231.

\renewcommand\appendixname{Supplementary}
\pagebreak
\onecolumngrid
\appendix
\pagebreak

\chapter{\Large \bf Supplementary Material}
\vspace{1cm}

\renewcommand\thesection{Note \arabic{section}}
\renewcommand\thefigure{\textbf{\arabic{figure}}}
\renewcommand\thetable{\textbf{\arabic{table}}}
\renewcommand{\figurename}{\textbf{Supplementary Figure}}
\renewcommand{\tablename}{\textbf{Supplementary Table}}
\renewcommand\theequation{S\arabic{equation}}

\newcommand{\IRone}{\Gamma_1}
\newcommand{\IRtwo}{\Gamma_2}
\newcommand{\IRthree}{\Gamma_3}
\newcommand{\IRfour}{\Gamma_4}
\newcommand{\IRfive}{\Gamma_5}
\newcommand{\IRsix}{\Gamma_6}
\renewcommand{\IRseven}{\Gamma_7}
\newcommand{\IReight}{\Gamma_8}
\newcommand{\IRnine}{\Gamma_9}
\renewcommand{\IRten}{\Gamma_{10}}
\renewcommand{\IReleven}{\Gamma_{11}}
\newcommand{\IRtwelve}{\Gamma_{12}}

\renewcommand{\r}{\mathbf{r}}

\setcounter{figure}{0}    
\setcounter{table}{0}  
\setcounter{equation}{0}      

\begin{figure}[h]
	\centering
	\includegraphics[width=16cm]{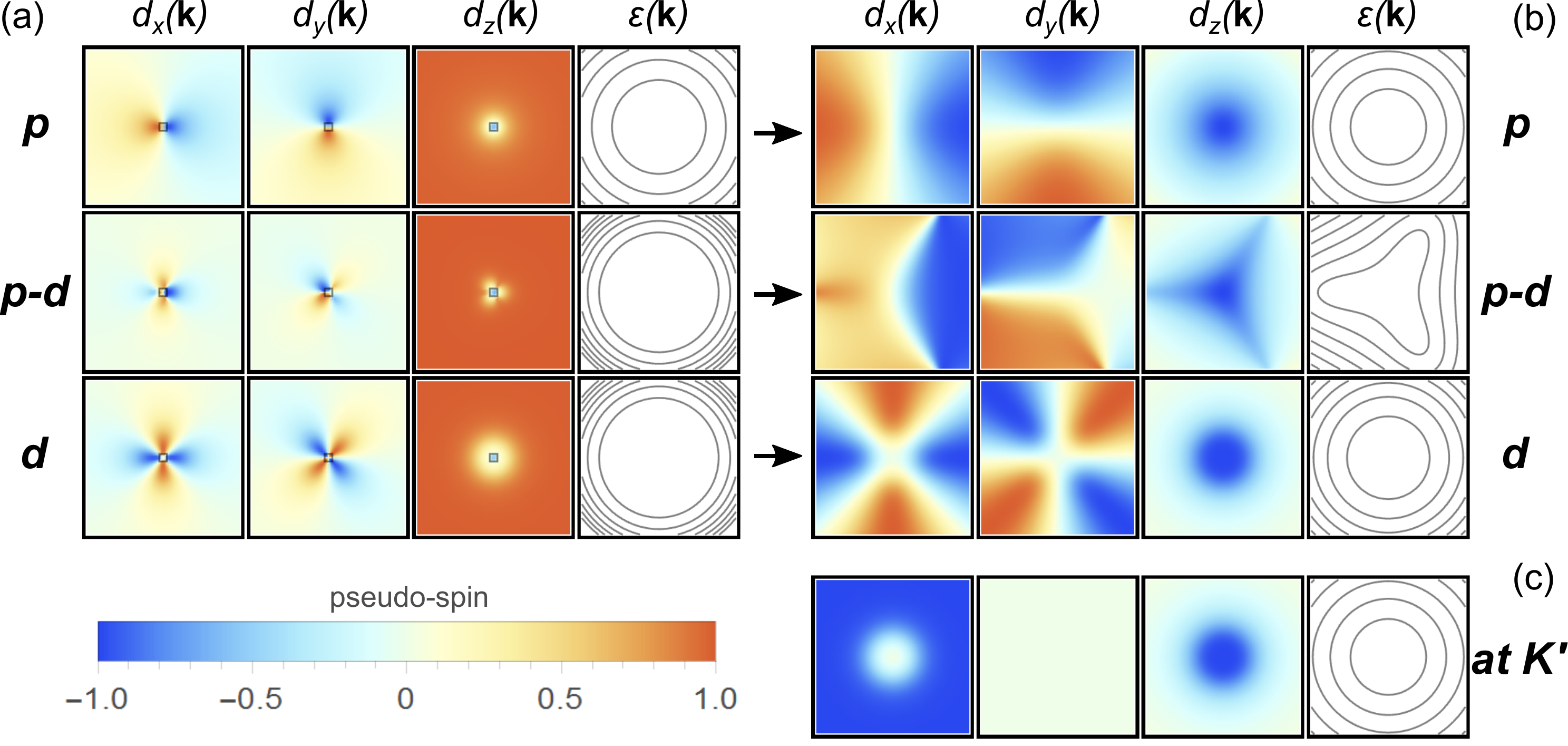}
	\caption{\textbf{Floquet-Bloch pseudospin textures at $\K$ and $\KP$.} (a) At $\K$ the pseudospin texture interpolates between $p$-wave ($v_p\neq 0 , v_d = 0$) and $d$-wave ($v_p= 0 , v_d \neq 0$)  winding upon increase of trigonal distortion. Given \textit{a priori} knowledge that the band structure remains gapped globally, a topological invariant can be assigned by introducing a regularizer $\eta$ that ensures that the pseudospin points into the $z$ direction for $k\to\infty$ and the $\k$-space manifold can thus be compactified to a sphere. Trigonal distortion is seen clearly when zooming into the vicinity (black squares) of the $\K$ point,  shown enlarged in (b). Here, while the local $p$-$d$ wave texture appears distorted, globally either $p$-wave or $d$-wave winding is recovered as seen in (a). At $\KP$, the pseudospin is trivial, with a vanishing $y$-component, depicted in (c). }
\end{figure}


\begin{table}[h]
	\centering
	\caption{\textbf{The single group $C_{\textrm{3h}}$}}
	\vspace{0.2cm}
	\subfigure[~$C_{\textrm{3h}}$ single group character table, with $\varOmega = \exp(2\pi i/3)$.]{
	\hspace{1.5cm} \begin{tabular}{| r | c  c  c  c  c  c | c |}
		\hline
		&  $E$  &  $C_3^+$  &  $C_3^-$  &  $\sigma_h$  &  $S_3^+$  &  $S_3^-$  &  invariants \\
		\hline
		$A'$       	& 1	& 1		& 1		& 1	& 1		& 1		& 				$x^2+y^2$, $z^2$ \\
		$E'$        	& 1	& $\eeA$	& $\eeC$	& 1	& $\eeA$	& $\eeC$	& $x-iy$, $(x+iy)^2$\\
		$\bar{E}'$ 	& 1	& $\eeC$	& $\eeA$	& 1	& $\eeC$	& $\eeA$	& $x+iy$, $(x-iy)^2$ \\
		$A''$         	& 1	& 1		& 1		& -1	& -1		& -1		& $z$			\\
		$E''$     		& 1	& $\eeA$	& $\eeC$	& -1	& -$\eeA$	& -$\eeC$	&				 $(x-iy)z$ \\
		$\bar{E}''$	& 1	& $\eeC$	& $\eeA$	& -1	& -$\eeC$	& -$\eeA$	&				 $(x+iy)z$ \\
		\hline
	\end{tabular}
	}
	
	\subfigure[~Selection rules for electric dipole transitions, \newline for circular polarization]{
\hspace{3.5cm}	\begin{tabular}{ c | c  c  c |}

	     	  &  $A'$  &  $E'$  &  $\bar{E}'$ \\
		\hline
	  $A'$  &           &    $\circlearrowleft$     &    $\circlearrowright$            \\
 	  $E'$  &    $\circlearrowright$       &         &     $\circlearrowleft$           \\
	  $\bar{E}'$  &   $\circlearrowleft$        &  $\circlearrowright$       &                \\
	  \hline
	\end{tabular} \hspace{3cm}
	}
\end{table}

\begin{table}[t]
	\centering
	\caption{\textbf{The double group $\bar{C}_{\rm 3h}$}}
	\vspace{0.2cm}
	\subfigure[~$\bar{C}_{\rm 3h}$ double group character table, with complex characters $\varOmega = \exp(2\pi i/3)$.]{
	\begin{tabular}{| r | c  c  c  c  c  c  c  c  c  c  c  c | c |}
		\hline
 & $E$ & $C_3^+$ & $C_3^-$ & $\sigma_h$ & $S_3^+$ & $S_3^-$ & $\bar{E}$ & $\bar{C}_3^+$ & $\bar{C}_3^-$ & $\bar{\sigma}_h$ & $\bar{S}_3^+$ & $\bar{S}_3^-$ & invariants \\
 \hline
 $\IRone$ & $1$ & $1$ & $1$ & $1$ & $1$ & $1$ & $1$ & $1$ & $1$ & $1$ & $1$ & $1$  &    $x^2+y^2, z^2$ \\
 $\IRtwo$ & $1$ & $\eeA$ & $\eeC$ & $1$ & $\eeA$ & $\eeC$ & $1$ & $\eeA$ & $\eeC$ & $1$ & $\eeA$ & $\eeC$ &     $x-iy, (x+iy)^2$ \\
 $\IRthree$ & $1$ & $\eeC$ & $\eeA$ & $1$ & $\eeC$ & $\eeA$ & $1$ & $\eeC$ & $\eeA$ & $1$ & $\eeC$ & $\eeA$ &     $x+iy, (x-iy)^2$ \\
 $\IRfour$ & $1$ & $1$ & $1$ & $-1$ & $-1$ & $-1$ & $1$ & $1$ & $1$ & $-1$ & $-1$ & $-1$  &      $z$ \\
 $\IRfive$ & $1$ & $\eeA$ & $\eeC$ & $-1$ & $-\eeA$ & $-\eeC$ & $1$ & $\eeA$ & $\eeC$ & $-1$ & $-\eeA$ & $-\eeC$  &     $(x-i y)z$ \\
 $\IRsix$ & $1$ & $\eeC$ & $\eeA$ & $-1$ & $-\eeC$ & $-\eeA$ & $1$ & $\eeC$ & $\eeA$ & $-1$ & $-\eeC$ & $-\eeA$  &     $(x+i y)z$ \\
 $\IRseven$ & $1$ & $-\eeA$ & $-\eeC$ & $i$ & $-i\eeA$ & $i\eeC$ & $-1$ & $\eeA$ & $\eeC$ & $-i$ & $i\eeA$ & $-i\eeC$  &     $\uparrow_z$ \\
 $\IReight$ & $1$ & $-\eeC$ & $-\eeA$ & $-i$ & $i\eeC$ & $-i\eeA$ & $-1$ & $\eeC$ & $\eeA$ & $i$ & $-i\eeC$ & $i\eeA$  &     $\downarrow_z$ \\
 $\IRnine$ & $1$ & $-\eeA$ & $-\eeC$ & $-i$ & $i\eeA$ & $-i\eeC$ & $-1$ & $\eeA$ & $\eeC$ & $i$ & $-i\eeA$ & $i\eeC$ & \\
 $\IRten$ & $1$ & $-\eeC$ & $-\eeA$ & $i$ & $-i\eeC$ & $i\eeA$ & $-1$ & $\eeC$ & $\eeA$ & $-i$ & $i\eeC$ & $-i\eeA$ & \\
 $\IReleven$ & $1$ & $-1$ & $-1$ & $i$ & $-i$ & $i$ & $-1$ & $1$ & $1$ & $-i$ & $i$ & $-i$ & \\
 $\IRtwelve$ & $1$ & $-1$ & $-1$ & $-i$ & $i$ & $-i$ & $-1$ & $1$ & $1$ & $i$ & $-i$ & $i$ & \\
 \hline
 	\end{tabular}
 	}
 	
 	\subfigure[~single-group and double-group irreducible representations of the Wannier orbital basis]{
 	\hspace{2cm} \begin{tabular}{| c | c  | c |}
 	\hline
 	state  &  single group IR  &  double group IR \\
 	\hline
 	$\ket{d_{3z^2-r^2},\uparrow}$   &  $A'$   &   $\IRseven$  \\
	$\ket{d_{x^2-y^2}-id_{2xy},\uparrow}$    &  $E'$  &   $\IRten$     \\
	$\ket{d_{x^2-y^2}+id_{2ixy},\uparrow}$   &  $\bar{E}'$   &   $\IReleven$    \\
	$\ket{d_{xz-iyz},\uparrow}$    &  $E''$  &   $\IReight$   \\
	$\ket{d_{xz+iyz},\uparrow}$    &  $\bar{E}''$  &   $\IRtwelve$    \\
	\hline
 	$\ket{d_{3z^2-r^2},\downarrow}$   &  $A'$  &   $\IReight$   \\
	$\ket{d_{x^2-y^2}-id_{2xy},\downarrow}$    &  $E'$  &   $\IRtwelve$  \\
	$\ket{d_{x^2-y^2}+id_{2xy},\downarrow}$    &  $\bar{E}'$  &  $\IRnine$    \\
	$\ket{d_{xz-iyz},\downarrow}$   &  $E''$  &    $\IReleven$   \\
	$\ket{d_{xz+iyz},\downarrow}$     & $\bar{E}''$   &   $\IRseven$   \\
	\hline
 	\end{tabular} \hspace{2cm}
	}
	\subfigure[~optical selection rules]{
		\begin{tabular}{ c c | c  c  c  | c  c  c |}
		\multicolumn{2}{c}{\multirow{2}{*}{ $\bar{C}_{\rm 3h}$ IR }}	& \multicolumn{3}{c}{$\boldsymbol{\Uparrow}$}\vspace{0.1cm} & \multicolumn{3}{c}{$\boldsymbol{\Downarrow}$} \\
	    & 	  &  $\IRseven$  &  $\IRten$  &  $\IReleven$  &  $\IReight$  &  $\IRtwelve$  &  $\IRnine$ \\
		\hline
	& $\IRseven$  &           &    $\circlearrowleft$     &    $\circlearrowright$     & & &        \\
 	$\boldsymbol{\Uparrow}$ &  $\IRten$  &    $\circlearrowright$       &         &     $\circlearrowleft$      & & &      \\
	&  $\IReleven$  &   $\circlearrowleft$        &  $\circlearrowright$       &          & & &       \\
	  \hline
	& $\IReight$  & & & &           &    $\circlearrowleft$     &    $\circlearrowright$            \\
 	$\boldsymbol{\Downarrow}$ & $\IRtwelve$ & & &  &    $\circlearrowright$       &         &     $\circlearrowleft$           \\
	&  $\IRnine$  & & & &   $\circlearrowleft$        &  $\circlearrowright$       &                \\
	  \hline
	\end{tabular}
	}
\end{table}

\begin{figure}[h]
\centering
\includegraphics[width=16cm]{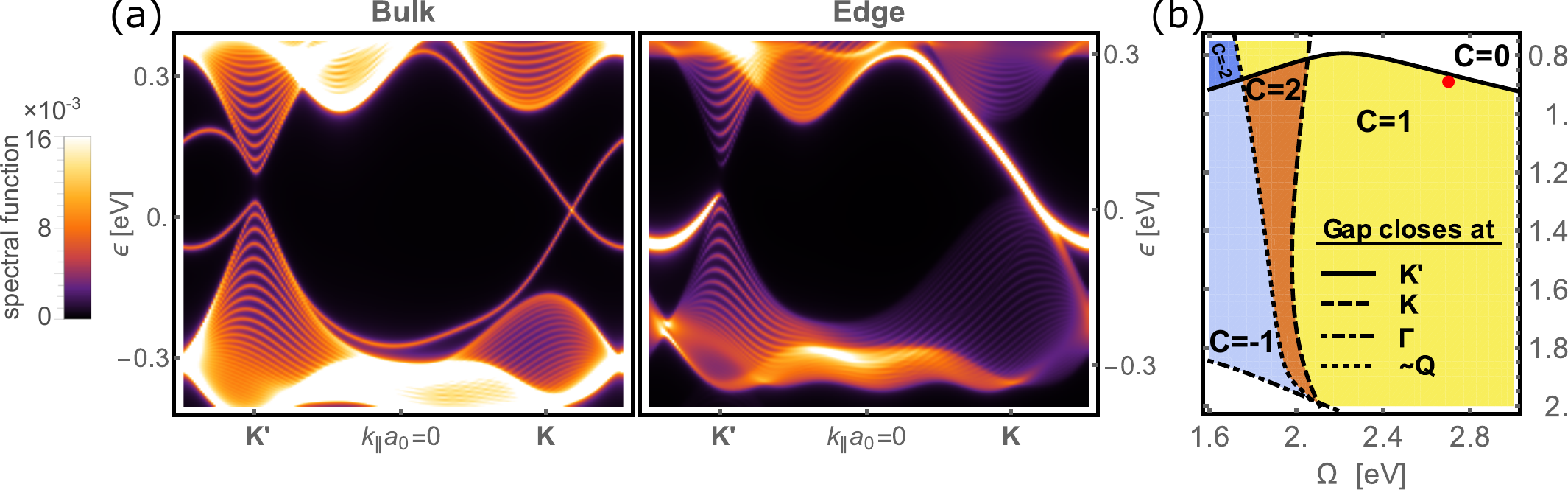}
\caption{\textbf{Photo-induced inversions in WS$_2$ for strong pump fields.} (a) A sufficiently blue-detuned pump field closes the equilibrium gap selectively at a single valley $\KP$, transitioning into a $\Ch=1$ phase. One of the two trivial equilibrium edge states disappears, leaving a single chiral edge mode that spans the band gap. The corresponding phase diagram of Floquet Chern numbers at strong pump strengths is depicted in (b), with (a) corresponding to parameters $A=0.9, \omega=2.7eV$. Further increase of the pump amplitude or decrease of frequency triggers additional gap closings at $\K$, $\boldsymbol{\Gamma}$, and around the second conduction band minimum $\mathbf{Q}$, inducing a mosaic of possible Chern numbers for the photo-modulated valence band. Closing the gap at $\K$, $\KP$ changes the Chern number by $\pm 1$, whereas $C_3$ symmetry dictates closing the gap at $\mathbf{Q}$ must happen at three points in the Brillouin zone, triggering a change of the Chern number by $\pm 3$. Progressive flattening of the valence band dispersion at strong pump fields indicates the onset of Wannier-Stark physics.}
\end{figure}

\clearpage
\section{Symmetry Analysis and the Role of Spin-Orbit Coupling}

In the absence of spin-orbit coupling (SOC), the relevant bands near $\K,\KP$ can be classified according to the single-group irreducible representations (IRs) of the point group $C_{\textrm{3h}}$, denoted $A',E',\bar{E}',E'',\bar{E}''$ \cite{Sbromley1972band,Smattheis1973band}. The corresponding character table and relevant invariants are depicted in Supplementary Table 2. As discussed in the main text, reflection symmetry $\sigma_h$ guarantees that bands $A',E',\bar{E}'$ remain decoupled from $E'',\bar{E}'$ across the entire Brillouin zone \cite{Sxiao2012coupled,Szahid2013generic,Scappelluti2013tight,Srostami2013effective,Sliu2013three}, and the analysis can therefore be constrained to three relevant bands $A',E',\bar{E}'$ only. In 2H monolayer TMDCs, the conduction band transforms as $A'$ and is dominantly composed of the transition metal $d_{3z^2-r^2}$ orbital, whereas the valence and relevant higher-energy band transform as $E',\bar{E}'$ with a dominant contribution of $d_{x^2-y^2} \pm i d_{2xy}$ orbitals. Furthermore, spin z is a good quantum number.

In equilibrium, the band structure can be expanded in $\k$ around $\K,\KP$ by starting from Hamiltonian
\begin{align}
	\Ham &= \Ham_0 + \Ham_{\rm SOC} + \Ham_\k
\end{align}
where
\begin{align}
	\Ham_0 &= \frac{1}{2m_0} \mom^2 + V(\r) \\
	\Ham_{\rm SOC} &= \frac{\hbar}{4m_0^2c_0^2} \mom \cdot \hat{\boldsymbol{\sigma}} \times \boldsymbol{\nabla}V(\r) \\
	\Ham_\k &= \frac{\hbar^2\k^2}{2m_0} + \frac{\hbar}{2m_0} \k\cdot\left[ \mom +  \frac{\hbar}{4m_0^2c_0^2} \hat{\boldsymbol{\sigma}} \times \boldsymbol{\nabla}V(\r) \right]
\end{align}
Here, $\hat{\boldsymbol{\sigma}}$ are the Pauli matrices, and $V(\r)$ is the crystal potential.

The role of SOC can now be understood in two complementary ways, by either considering the eigenstates of $\Ham_0$ as IRs of the $C_{\textrm{3h}}$ single group and treating SOC as a perturbation, or by starting from the true Bloch eigenstates of $\Ham_0 + \Ham_{\rm SOC}$ at $\K,\KP$, as IRs of the $\bar{C}_{\rm 3h}$ double group. 
  In the former case, the eigenstates of $\Ham_0$ are spin-z eigenstates with a given single-group IR, namely $\ket{A'[d_{3z^2-r^2}],\sigma}$, $\ket{E'[d_{x^2-y^2}+id_{2xy}],\sigma}$ and $\ket{\bar{E}'[d_{x^2-y^2}-id_{2xy}],\sigma}$ with $\sigma=\uparrow_z,\downarrow_z$. To understand the effect of SOC, it is useful to decompose $\Ham_{\rm SOC}, \Ham'_\k$:
\begin{align}
	\Ham_{\rm SOC}^z &= \hat{\sigma}_z \left[ \hat{p}_x \partial_y - \hat{p}_y \partial_x \right] V(\r) \\
	\Ham_{\rm SOC}^{\uparrow\downarrow} &= 2 i \hat{\sigma}^{-} \left[ \hat{p}_z \partial_+ V(\r) - \hat{p}_+ \partial_z V(\r) \right] + \hc \\
	\Ham_\k^z &= \frac{\hbar^2\k^2}{2m_0} + \frac{\hbar}{2m_0} \left[ k_+ \cdot \left( \mom - i  \hat{\sigma}_z  \frac{\hbar}{4m_0^2c_0^2} \partial_- V(\r) \right) + \hc \right] \\
	\Ham_\k^{\uparrow\downarrow} &= \frac{\hbar}{2m_0} k_+ \cdot \left( \mom + i \hat{\sigma}_-  \frac{\hbar}{4m_0^2c_0^2} \partial_z V(\r) \right) + \hc 
\end{align}
Here, $\Ham_{\rm SOC}^z$ transforms as $A'$ and acts as a mere Zeeman shift, whereas the spin-flip contribution $\Ham_{\rm SOC}^{\uparrow\downarrow}$ transforms as $E'',\bar{E}''$ and hence couples states with opposite parity under $\sigma_h$. The latter entails a mixing between conduction band state $\ket{A',\sigma}$ and $\ket{E''(\bar{E}''),-\sigma}$ as well as between the higher-energy conduction band $\ket{\bar{E}',\uparrow}$ and $\ket{E'',\downarrow}$ while leaving the opposite spin $\ket{\bar{E}',\downarrow}$ unmixed. This mixing of $\uparrow_z,\downarrow_z$ states is however suppressed since the energy difference between the $A',E'$ and $E''$ bands is larger than the spin-orbit coupling, and the SOC at $\K,\KP$ can be well-captured as an effective Zeeman shift while approximately leaving spin as a good quantum number \cite{Scappelluti2013tight,Sliu2013three,Sgibertini2014spin,Skormanyos2013monolayer,Sfeng2012intrinsic}.
Away from $\K,\KP$, $\Ham_\k^z$ transforms as $E',\bar{E}'$ and imparts an additional momentum-dependent Zeeman shift. Conversely, $\Ham_\k^{\uparrow\downarrow}$ transforms as $A''$, which does not couple the $A'$ states of the original conduction band while weakly mixing the $E',E''$ bands.

Consider now the double group, with its character table given in Supplementary Table 2. Spin-flip mixing with the $E'',\bar{E}''$ bands necessarily reduces the number of band IRs from $5$ IRs $\times 2$ spin orientations, to $6$ double-group IRs. These again decompose into two manifolds denoted $\Uparrow,\Downarrow$ that remain decoupled over the entire Brillouin zone. The identification of band states with single group and double group IRs is given in Supplementary Table 2(b). Note that the states listed are not an eigenbasis of $\Ham_0 + \Ham_{\rm SOC}$: instead, the true eigenbasis will be a superposition of states of equal double-group IR, governed by the strength of SOC. At $\K,\KP$, one can immediately deduce that the eigenbasis entails mixing of $\ket{d_{3z^2-r^2},\sigma}$ and $\ket{d_{xz+\sigma\cdot iyz},-\sigma}$, of $\ket{d_{x^2-y^2+2ixy}+id_{2xy},\uparrow}$ and $\ket{d_{xz-iyz},\downarrow}$, and of$\ket{d_{xz+iyz},\uparrow}$ and $\ket{d_{x^2-y^2}-id_{2ixy},\downarrow}$, while leaving states $\ket{d_{x^2-y^2}-id_{2xy},\uparrow}$ and $\ket{d_{x^2-y^2}+id_{2xy},\downarrow}$ unmixed and as proper eigenstates of $\Ham_0 + \Ham_{\rm SOC}$.

Here, the $\mathbf{k}.\mathbf{p}$ perturbation transforms as $\IRtwo, \IRthree$ and couples the three $\Uparrow$ IRs $\IRseven,\IRten,\IReleven$ ($\Downarrow$ IRs $\IReight,\IRnine,\IRtwelve$) in an equivalent manner as in the single-group case of $A',E',\bar{E}'$ without spin-orbit coupling, or with SOC but without spin-flip terms. Consequently, the selection rules of electric dipole transitions for circular polarization are exactly equivalent, as shown in Supplementary Table 2(c). For this reason, we chose to simply adopt the single-group notation and label the spin manifolds as $\uparrow_z,\downarrow_z$, while keeping in mind that SOC is indeed significant for certain TMDCs and enters through a one-to-one correspondence with the double group IRs and spin manifolds $\Uparrow,\Downarrow$.

\section{Floquet $k.p$ Theory}

A central result of the main text, the effective Floquet $\kp$ Hamiltonians (1) and (2), as determined from symmetry considerations of the non-equilibrium problem, characterize the photo-induced band inversions at $\K$ and $\KP$ and describe transitions between zero, one and two chiral edge modes as a function of pump strength and frequency. To further shed light on the procedure of Floquet $\kp$ theory as outlined in the Methods section, we consider here an explicit derivation of Eqns. (1) and (2) of the main text, in the limit of small $A$.

We set $\hbar = 1$. The first step concerns finding the time-dependent Floquet eigenbasis directly at $\K,\KP$, as eigenfunctions of $\Ham_{\textrm{eq}} + \Ham_{\textrm{pump}}(t)$ [Eq. (7), (8) in the Methods section]. Working with $C_{\textrm{3h}}$ IRs, this equivalently amounts to solving the time-independent Floquet eigenbasis of the Floquet Hamiltonian:
\begin{align}
	\Ham_{0F} = \sum_{mn\alpha} & \left(\E_{n\alpha} + m\Omega + \frac{e^2A^2}{2m_0} \right) \ketbra{m;\alpha n}{m;\alpha n} \notag\\
	+~ e A\sum_{mnn'} & \left(g_{nn'}^{A'E'} \ketbra{m+1;A',n}{m;E',n'} \right. \notag\\
	+& ~ g_{nn'}^{\bar{E}'A'} \ketbra{m+1;\bar{E}',n}{m;A',n'} \notag\\
	+& \left. g_{nn'}^{E'\bar{E}'} \ketbra{m+1;E',n}{m;\bar{E}',n'} + \hc  \right)
	\label{eq:Floquet0}
\end{align}
Here, $m$ is the Floquet index; $n,\alpha$ index the n\textsuperscript{th} band in the $C_{\textrm{3h}}$ IRs $\alpha=A',E',\bar{E}'$, and $g_{nn'}^{\alpha\alpha'} = \frac{1}{2m_0}\braOPket{\alpha,n}{\left(\SOCmom_x - i \SOCmom_y\right)}{\alpha',n'}$ are the momentum matrix elements for allowed dipole transitions that are obtained from \textit{ab initio} calculations.

Having determined the new eigenbasis of (\ref{eq:Floquet0}) that admixes different Floquet side bands and IRs of $C_{\textrm{3h}}$, deviations in momentum away from $\K,\KP$ can now be treated as a perturbation in $\Ham'(t)$ [Eq. (9) in the Methods section]:
\begin{align}
	\Ham' = k_+ ~&\left\{ \sum_{mn\alpha} \frac{e}{m_0} \frac{A}{2} \ketbra{m-1;\alpha,n}{m;\alpha,n} \right. \notag\\
	&~~+ \sum_{mnn'} \left( g_{nn'}^{A'E'} \ketbra{m;A',n}{m;E',n'}  \right. \notag\\
	&~~~~~~~~+ ~g_{nn'}^{\bar{E}'A'} \ketbra{m;\bar{E}',n}{m;A',n'} \notag\\
	&~~~~~~~~+ \left. \left. g_{nn'}^{E'\bar{E}'} \ketbra{m;E',n}{m;\bar{E}',n'} \right) \vphantom{\sum_{mn\alpha} \frac{A}{2}} \right\} + \hc
	\label{eq:FloquetKP}
\end{align}

At $\K$, and in the limit of small $A$, the Floquet eigenstates of $\Ham_{0F}$ that compose the eigenbasis of Eq. (1) of the main text can be found perturbatively:
\begin{align}
	\ket{\Psi_{\K,1}} &= \ket{0; A',{\rm CB}} + eA \left[\sum_n \frac{(g_{{\rm CB},n}^{A'E'})^\star  ~\ket{-1; E',n}  }{\E_{A',{\rm CB}} - \E_{E',n} + \Omega} + \sum_n \frac{g_{n,{\rm CB}}^{\bar{E}'A'}  ~\ket{+1; \bar{E}',n}  }{\E_{A',{\rm CB}} - \E_{\bar{E}',n} - \Omega} \right] + \mathcal{O}(A^2) \\
	\ket{\Psi_{\K,2}} &= \ket{-1; \bar{E}',{\rm XB}} + eA \left[\sum_n \frac{(g_{{\rm {\rm XB}},n}^{\bar{E}'A'})^\star ~\ket{-2; A',n}}{\E_{\bar{E}',{\rm XB}} - \E_{A',n} + \Omega}  + \sum_n \frac{g_{n,{\rm XB}}^{E'\bar{E}'} ~\ket{0; E',n} }{\E_{\bar{E}',{\rm XB}} - \E_{E',n} - \Omega}  \right] + \mathcal{O}(A^2)
\end{align}
In a second step, one can now start from this eigenbasis and perturb in $\k$, which amounts to a perturbation in $\Ham'$ [Eqns. (9), (11) in the Methods section]. Here, a coupling between $\ket{\Psi_1}$, $\ket{\Psi_2}$ is mediated via the $m=0; E'$ VB (and via all other bands of $E'$ IR) to linear order $\sim A k_-$, as well as via the $m=-1;A'$ CB or the $m=0;\bar{E}'$ XB to quadratic order $\sim A k_+^2$. To second order in $\k$, we arrive at the effective Hamiltonian (1) quoted in the main text, where, to linear order in $A$, we obtain $p$- and $d$-wave couplings of the form
\begin{align}
	v_p &= eA ~\sum_n g_{{\rm CB},n}^{A'E'} ~g_{n,{\rm XB}}^{E'\bar{E}'} ~ \left( \frac{1}{\E_{A',{\rm CB}} - \E_{E',n} + \Omega}  + \frac{1}{\E_{\bar{E}',{\rm XB}} - \E_{E',n} - \Omega} \right) + \mathcal{O}(A^2)\\
\
	v_d &= eA \left\{ \frac{1}{4m_0}~ (g_{{\rm {\rm XB}},{\rm CB}}^{\bar{E}'A'})^\star \left( \frac{1}{\E_{A',{\rm CB}} - \E_{\bar{E}',{\rm XB}}} - \frac{1}{\Omega} + \frac{1}{\Omega} + \frac{1}{\E_{\bar{E}',{\rm XB}} - \E_{A',{\rm CB}}} \right) + \right. \notag\\
	&+ \frac{1}{2} \sum_{nn'} \frac{  g_{n,{\rm XB}}^{E'\bar{E}'}  \left( g_{nn'}^{E'\bar{E}'} g_{n',{\rm CB}}^{\bar{E}'A'} \right)^\star }{\E_{\bar{E}',{\rm XB}} - \E_{E',n} - \Omega} \left( \frac{1}{\E_{A',{\rm CB}} - \E_{\bar{E}',n'}} + \frac{1}{\E_{E',n} - \E_{\bar{E}',n'}} \right)    + \notag\\
	&+ \frac{1}{2} \left. \sum_{nn'} \frac{ g_{{\rm CB},n'}^{A'E'} \left( g_{{\rm {\rm XB}},n}^{\bar{E}'A'} g_{nn'}^{A'E'}  \right)^\star }{\E_{A',{\rm CB}} - \E_{E',n'} + \Omega} \left( \frac{1}{\E_{E',n'} - \E_{A',n}} + \frac{1}{\E_{\bar{E}',{\rm XB}} - \E_{A',n} } \right) \right\}  + \mathcal{O}(A^2)
\end{align}
together with Dirac and band mass terms
\begin{align}
	M &= \E_{\bar{E}',{\rm XB}} - \E_{A',{\rm CB}} - \Omega + \underbrace{\mathcal{O}(A^2)}_{\textrm{Stark shift}} \\
\
	B &= \sum_{n\gamma} \left[ \frac{ \left| g_{n,{\rm CB}}^{\gamma,A'} \right|^2 }{\E_{A',{\rm CB}} - \E_{n,\gamma}} - \frac{ \left| g_{n,{\rm XB}}^{\gamma,\bar{E}'} \right|^2 }{\E_{\bar{E}',{\rm XB}} - \E_{n,\gamma}} \right] + \mathcal{O}(A^2)
\end{align}

As can be seen from the above derivations, an essential consequence of the minimally three-band description of electron-photon coupling in TMDCs is the appearance of the linear ``Dirac'' coupling $v_p$ on the level of the effective Floquet Hamiltonian for $\K$. Indeed, $v_p$ necessarily depends on dipole matrix elements that couple all three bands $\sim g_{{\rm CB},n}^{A'E'} ~g_{n,{\rm XB}}^{E'\bar{E}'}$ with $n$ the VB or further bands of $E'$ IR. Therefore, besides naturally precluding the appearance of a red-detuned regime, equilibrium two-band toy models of TMDCs can never capture a non-equilibrium transition between $C=2$ and $C=1$ in the blue-detuned regime.

Conversely, at $\KP$, the perturbative Floquet eigenstates of $\Ham_{0F}$ that compose the eigenbasis of Eq. (2) in the main text read:
\begin{align}
	\ket{\Psi_{\KP,1}} &= \ket{0; A',{\rm CB}} + eA \left[ \sum_n \frac{(g_{{\rm CB},n}^{A'E'})^\star  ~\ket{-1; E',n} }{\E_{A',{\rm CB}} - \E_{E',n} + \Omega}  + \sum_n \frac{g_{n,{\rm CB}}^{\bar{E}'A'}  ~\ket{+1; \bar{E}',n}  }{\E_{A',{\rm CB}} - \E_{\bar{E}',n} - \Omega} \right] + \mathcal{O}(A^2) \\
	\ket{\Psi_{\KP,2}} &= \ket{-1; E',{\rm XB}} + eA \left[ \sum_n \frac{(g_{{\rm {\rm XB}},n}^{E'\bar{E}'})^\star  ~ \ket{-2; \bar{E}',n} }{\E_{E',{\rm XB}} - \E_{\bar{E}',n} + \Omega} + \sum_n \frac{ g_{n,{\rm XB}}^{A'E'}  ~\ket{0; A',n}   }{\E_{E',{\rm XB}} - \E_{A',n} - \Omega} \right] + \mathcal{O}(A^2)
\end{align}
Here,  $\ket{\Psi_{\KP,1}}$ and $\ket{\Psi_{\KP,2}}$ remain decoupled to linear order in $\k$. Instead, we arrive at a quadratic coupling of the form $\sim Ak_+k_- = A\k^2$ as discussed in the main text, with
\begin{align}
	v' &=eA g_{{\rm CB},{\rm XB}}^{A',E'} \frac{1}{4m_0} \left( \frac{1}{\E_{A',{\rm CB}} - \E_{E',{\rm XB}}} - \frac{1}{\Omega} + \frac{1}{\Omega} + \frac{1}{\E_{E',{\rm XB}} - \E_{A',{\rm CB}}} \right) + \notag\\
	&+ \sum_{nn'} eA \frac{g_{n',{\rm XB}}^{A'E'} }{\E_{E',{\rm XB}} - \E_{A',n'} - \Omega} \left[ g_{n,n'}^{\bar{E}'A'} (g_{n,{\rm CB}}^{\bar{E}'A'})^\star  \left( \frac{1}{\E_{A',{\rm CB}} - \E_{\bar{E}',n}} + \frac{1}{\E_{A',n} - \E_{\bar{E}',n'}} \right) \right. + \notag\\
	& \left. ~~~~~~~~~~+ (g_{n'n}^{A'E'})^\star g_{{\rm CB},n}^{A'E'} \left(  \frac{1}{\E_{A'{\rm CB}} - \E_{E',n}} + \frac{1}{\E_{A',n'} - \E_{E',n}}  \right) \right] + \notag\\
	& + \sum_{nn'} eA \frac{ g_{{\rm CB},n}^{A'E'} }{\E_{A',{\rm CB}} - \E_{E',n} + \Omega} \left[ g_{n',{\rm XB}}^{A'E'} (g_{n',n}^{A'E'})^\star \left( \frac{1}{\E_{E',n} - \E_{A',n'}} + \frac{1}{\E_{E',{\rm XB}} - \E_{A',n'}} \right) +  \right. \notag\\
	& \left. ~~~~~~~~~~+ (g_{{\rm {\rm XB}},n'}^{E'\bar{E}'})^\star g_{n,n'}^{E'\bar{E}'} \left( \frac{1}{\E_{E',n} - \E_{\bar{E}',n'}} + \frac{1}{\E_{E',{\rm XB}} - \E_{\bar{E}',n'}} \right)  \right]
\end{align}

\section{Topological Classification and Pseudospin Textures}

The main text classifies photo-induced topological phase transitions via local effective Floquet $\mathbf{k}.\mathbf{p}$ Hamiltonians at $\K,\KP$. The key idea is to understand the global topology via a local classification of band inversions at $\K,\KP$, which relies on \textit{a-priori} knowledge that 1) the Floquet spectrum is gapped globally and 2) the Berry curvature behaves benign at other high-symmetry points in the Brillouin zone. Armed with this knowledge, a complementary view of local band inversions follows from considering so-called pseudospin textures around $\K$ and $\KP$. Starting from the local Floquet $\mathbf{k}.\mathbf{p}$ Hamiltonians $\Ham_\K(\k)$ and $\Ham_\KP(\k)$ of equations 1 and 2 of the main text, we can recast these in terms of pseudospin Pauli matrices $\boldsymbol{\sigma}$:
\begin{align}
	\Ham_{\nu=\K,\KP}(\k) = \E_0(\k) + \E(\k) ~\hat{\mathbf{d}}_\nu(\k) \cdot \boldsymbol{\sigma}  \label{eq:HamPseudo}
\end{align}
Here, $\E_0(\k) \pm \E(\k)$ is the dispersion of $\Ham_\K(\k)$ or $\Ham_\KP(\k)$, and $\hat{\mathbf{d}}_\nu(\k)$ is the pseudospin vector, with $|\hat{\mathbf{d}}_\nu(\k)| = 0$. Here, $\nu = \K,\KP$. The pseudospin vector equivalently follows from taking the expectation value $\left< \boldsymbol{\sigma} \right>$ of the Floquet-Bloch states around $\K,\KP$.

As discussed in the main text, the band inversion at $\KP$ is trivial, and the pseudospin obeys $d_\K^y(\k) = 0$. Conversely, at $\K$, the pseudospin becomes
\begin{align}
	\hat{\mathbf{d}}_\K(\k) = \frac{1}{N(\k)} \left[ v_p k_x + v_d (k_x^2 - k_y^2) , ~v_p k_y - 2 v_d k_x k_y , ~ M - B |\k|^2  \right]^\top
\end{align}
with a normalization $N(\k)$ to ensure that $|\hat{\mathbf{d}}_\K(\k)| = 1$. 

At $\K$, the effective Hamiltonian \ref{eq:HamPseudo} can be viewed as a $d$-wave generalization of the conventional massive Dirac Hamiltonian, with an additional band mass term in analogy to the $\mathbf{k}.\mathbf{p}$ model at $\Gamma$ for HgTe/CdTe quantum wells. Supplementary Figure 1 depicts the pseudospin textures upon increasing $v_d/v_p$ to enhance trigonal distortion. $v_p\neq0, v_d=0$ recovers the conventional massive Dirac fermion with a quadratic band mass term; here, the pseudospin has a $p$-wave winding around $\KP$ and the Chern number becomes $C=1$. In the opposite limit $v_p=0, v_d \neq 0$, the pseudospin acquires a $d$-wave winding around $\K$. Here, since the band mass term is quadratic only, the winding in $d^x(\k),d^y(\k)$ persists in principle to $k\to \infty$. Given the knowledge that the band structure is gapped globally and that the band inversion should be confined to high-symmetry points, this behavior is an artifact of the lower-order $\mathbf{k}.\mathbf{p}$ expansion.

More rigorously, quantization of the integral $\Ch = \frac{1}{2\pi} \int_{\mathbb{R}^2} d\k \mathcal{F}(\k)$ necessitates a compactification of k-space $\mathbb{R}^2$ to a non-contractible manifold. This can be motivated as follows: Since the Floquet $\k.\mathbf{p}$ theory can be expected to faithfully represent the physics only in the close vicinity around $\K$, momenta $\k$ far away from $\K$ should not affect $\Ch$. This can be enforced rigorously by adding to $\Ham$ an infinitesimal rotationally-symmetric regularizer $-\eta B' \hat{\sigma}_z |\k|^4$ with $\eta \to 0$, leading to:
\begin{align}
	\hat{\mathbf{d}}_\KP^{\textrm{(reg)}}(\k) = \frac{1}{N(\k)} \left[ v_p k_x + v_d (k_x^2 - k_y^2) , ~v_p k_y - 2 v_d k_x k_y , ~ M - B |\k|^2 (1 + \eta |\k|^2)  \right]^\top
\end{align}
It follows that the unit vector $\hat{\mathbf{d}}_\KP^{\textrm{(reg)}}(|\k| \to \infty) = -\textrm{sgn}(B) \hat{e}_z$ does not depend on the polar angle $\theta$ of $\k$, and $\mathbb{R}^2$ can be compactified to a sphere $S_2$ by identifying $\infty$ with the north pole without loss of information. 
The choice $\hat{\sigma}_z$ of the regularizer is motivated by noting that $|B| \gg |v_d|$ for any choice of pump field entails that $\hat{\mathbf{d}}(|\k| \to \infty) \sim -\textrm{sgn}(B) \hat{e}_z + \frac{v_d}{|B|} [ \cos(2\theta) \hat{e}_x - \sin(2\theta) \hat{e}_y ] $ already approximately points in the z-direction.

The band inversion can be seen clearly by looking at the behavior of $d^z(\k)$ close to $\K$, which switches sign when going from $\k=0$ to $\k\to\infty$. The intermediate regime of $p$-$d$-wave winding leads to a distorted pseudospin texture when looking at the close vicinity of $\K$, whereas a $d$-wave (or $p$-wave) texture is retained at large $\k$ to arrive at a $C=2$ or $C=1$ phase.

\section{Strong Pumping and Inversion of the Equilibrium Band Gap}

Discussions on photo-induced chiral edge states have focused so far solely on dynamically-generated gaps within the equilibrium conduction and valence bands, since a sizeable energy scale $\sim 1.5eV$ in WS$_2$ protects the equilibrium band gap from closing for weak pump fields. This picture changes conceivably when approaching the regime of Wannier-Stark physics at significantly higher pump strength. In the high-frequency limit, broken time-reversal symmetry then bestows an optical Stark shift of equal and opposite magnitude on $\K$ and $\KP$, that bridges the equilibrium band gap at a critical field strength $A$. The gap closes and reopens at $\KP$ to eliminate one branch of the trivial equilibrium edge states, leaving a single chiral edge state to bridge the Floquet-Bloch band gap at $\K$, as depicted in Supplementary Figure 2(a). Upon even further increase of $A$, the gap finally closes and reopens at $\Gamma$, returning to a trivial regime without chiral edge modes. Na\"ively, the flattening of the bands upon crossover to Wannier-Stark ladders at increasing pump strengths $A$ or decreasing frequencies $\Omega$ suggests that one should not expect to continue attributing special significance to the original high-symmetry points in this regime. Nevertheless, the system can undergo a series of gap closings confined to $\K, \KP, \Gamma$ upon further decrease of $\Omega$. The associated topological transitions change the Floquet Chern number by $\pm 1$, leading to a mosaic of photo-induced topological phases at high pump intensities. In addition, gap closures occur near the second conduction-band minimum $\mathbf{Q}$ in WS$_2$. As this is not a high-symmetry point, $C_3$ rotation symmetry dictates that the gap must instead close simultaneously at three distinct points in the Brillouin zone, changing the Chern number by $\pm 3$. We verified the corresponding phase diagram in Supplementary Figure 2(b) by numerically evaluating the Floquet Chern number.

\renewcommand\refname{Supplementary References}


\begin{thebibliography}{99}


\bibitem{Schmitt19092008}{
Schmitt, F. {\em et al.}
\newblock Transient electronic structure and melting of a charge density wave
  in TbTe$_3$.
\newblock {\em Science} 321, 1649 (2008).}

\bibitem{Fausti14012011}{
Fausti, D. {\em et al.}
\newblock Light-induced superconductivity in a stripe-ordered cuprate.
\newblock {\em Science} 331, 189 (2011).}

\bibitem{kim2012ultrafast}
{Kim, K.W. {\em et al.}
\newblock Ultrafast transient generation of spin-density-wave order in the
  normal state of BaFe$_2$As$_2$ driven by coherent lattice vibrations.
\newblock {\em Nature Mater.} 11, 497 (2012).}

\bibitem{mankowsky2014nonlinear}{
Mankowsky, R. {\em et al.}
\newblock Nonlinear lattice dynamics as a basis for enhanced superconductivity
  in YBa$_2$Cu$_3$O$_{6.5}$.
\newblock {\em Nature} 516, 71 (2014).}



\bibitem{lindner2011floquet}{
Lindner, N.H., Refael, G. \& Galitski, V.
\newblock Floquet topological insulator in semiconductor quantum wells.
\newblock {\em Nature Phys.}, 7, 490 (2011).}

\bibitem{wang2013observation}{
Wang, Y.H., Steinberg, H. Jarillo-Herrero, P. \& Gedik, N.
\newblock Observation of Floquet-Bloch states on the surface of a topological
  insulator.
\newblock {\em Science} 342, 453 (2013).}

\bibitem{mahmood2016floquetvolkov}{
Mahmood, F. {\em et al.}
\newblock Selective scattering between Floquet–Bloch and Volkov states in a topological insulator
\newblock {\em Nature Phys.}  12, 306 (2016). }

\bibitem{sie2014valley}{
Sie, E.J. {\em et al.}
\newblock Valley-selective optical stark effect in monolayer WS$_2$.
\newblock {\em Nature Mat.} 14, 290 (2014).}

\bibitem{kim2014ultrafast}
{Kim, J. {\em et al.}
\newblock Ultrafast generation of pseudo-magnetic field for valley excitons in
  WSe$_2$ monolayers.
\newblock {\em Science} 346, 1205 (2014).}







\bibitem{oka2009photovoltaic}{
Oka, T. \& Aoki, H.
\newblock Photovoltaic hall effect in graphene.
\newblock {\em Phys. Rev. B} 79, 081406 (2009).}


\bibitem{kitagawa2011transport}{
Kitagawa, T., Oka, T., Brataas, A., Fu, L. \& Demler, E.
\newblock Transport properties of nonequilibrium systems under the application
  of light: Photoinduced quantum Hall insulators without Landau levels.
\newblock {\em Phys. Rev. B} 84, 235108 (2011).}

\bibitem{sentef2015}{
Sentef, M.A., Claassen, M., Kemper, A.F., Moritz, B., Oka, T., Freericks, J.K. \& Devereaux, T.P.
\newblock Theory of Floquet band formation and local pseudospin textures in pump-probe photoemission of graphene.
\newblock {\em Nature Comm.} 6, 7047 (2015).}

\bibitem{xiao2012coupled}{
Xiao, D., Liu, G.-B., Feng, W., Xu, X. \& Yao, W.
\newblock Coupled spin and valley physics in monolayers of MoS$_2$ and other
  group-vi dichalcogenides.
\newblock {\em Phys. Rev. Lett.} 108, 196802 (2012).}


\bibitem{bromley1972band}{Bromley, R.A., Murray, R.B. \& Yoffe, A.D.
The band structures of some transition metal dichalcogenides. III. group VIA: Trigonal prism materials.
\newblock {\em J. of Physics C: Solid State Phys.} 5, 759 (1972).}

\bibitem{mattheis1973band}{
Mattheiss, L.F.
\newblock Band structures of transition-metal-dichalcogenide layer compounds.
\newblock {\em Phys. Rev. B} 8, 3719 (1973).}


\bibitem{feng2012intrinsic}{
Feng, W. {\em et al.}
\newblock Intrinsic spin hall effect in monolayers of group-VI dichalcogenides:
  A first-principles study.
\newblock {\em Phys. Rev. B} 86, 165108 (2012).}

\bibitem{shan2013spin}{
Shan, W.-Y., Lu, H.-Z. \& Xiao, D.
\newblock Spin hall effect in spin-valley coupled monolayers of transition
  metal dichalcogenides.
\newblock {\em Phys. Rev. B} 88, 125301 (2013).}



\bibitem{yu2014nonlinear}{
Yu, H., Wu, Y., Liu, G.-B., Xu, X. \& Yao, W.
\newblock Nonlinear valley and spin currents from fermi pocket anisotropy in 2D
  crystals.
\newblock {\em Phys. Rev. Lett.} 113, 156603 (2014).}

\bibitem{shan2015optical}{
Shan, W.-Y., Zhou, J. \& Xiao, D.
\newblock Optical generation and detection of pure valley current in monolayer
  transition-metal dichalcogenides.
\newblock {\em Phys. Rev. B} 91, 035402 (2015).}

\bibitem{muniz2015all}{
Muniz, R.A. \& Sipe, J.E.
\newblock All-optical injection of charge, spin, and valley currents in
  monolayer transition-metal dichalcogenides.
\newblock {\em Phys. Rev. B} 91, 085404 (2015).}


\bibitem{tahir2014photoinducedhall}{Tahir, M., Manchon, A. \& Schwingenschl\"ogl, U.
Photoinduced quantum spin and valley Hall effects, and orbital magnetization in monolayer MoS$_2$.
\newblock {\em Phys. Rev. B} 90, 125438 (2014).}


\bibitem{kogan2014energy}{
Kogan,E., Nazarov, V.U., Silkin, V.M. \& Kaveh, M.
\newblock Energy bands in graphene: Comparison between the tight-binding model
  and ab initio calculations.
\newblock {\em Phys. Rev. B} 89, 165430 (2014).}



\bibitem{zahid2013generic}{
Zahid, F., Liu, L., Zhu, Y., Wang, J. \& Guo, H.
\newblock A generic tight-binding model for monolayer, bilayer and bulk MoS$_2$.
\newblock {\em AIP Advances} 3, 052111 (2013).}

\bibitem{cappelluti2013tight}{Cappelluti, E., Rold{\'a}n, R., Silva-Guill{\'e}n, J.A., Ordej{\'o}n, P. \& Guinea, F.
\newblock Tight-binding model and direct-gap/indirect-gap transition in
  single-layer and multilayer MoS$_2$.
\newblock {\em Phys. Rev. B} 88, 075409 (2013).}

\bibitem{rostami2013effective}{
Rostami, H., Moghaddam, A.G. \& Asgari, R.
\newblock Effective lattice hamiltonian for monolayer MoS$_2$: Tailoring
  electronic structure with perpendicular electric and magnetic fields.
\newblock {\em Phys. Rev. B} 88, 085440 (2013).}

\bibitem{liu2013three}{
Liu, G.-B., Shan, W.-Y., Yao, Y., Yao, W. \& Xiao, D.
\newblock Three-band tight-binding model for monolayers of group-vib transition
  metal dichalcogenides.
\newblock {\em Phys. Rev. B} 88, 085433 (2013).}

\bibitem{winkler2003spin}{
Winkler, R., Papadakis, S.J., De Poortere, E.P. \& Shayegan, M.
\newblock {\em Spin-Orbit Coupling in Two-Dimensional Electron and Hole
  Systems}, volume~41.
\newblock Springer (2003).}


\bibitem{rudner2013anomalous}{
Rudner, M.S., Lindner, N.H., Berg, E. \& Levin, M.
\newblock Anomalous edge states and the bulk-edge correspondence for
  periodically driven two-dimensional systems.
\newblock {\em Phys. Rev. X} 3, 031005 (2013).}





\bibitem{liu2008quantum}{
Liu, C.-X., Qi, X.-L., Dai, X., Fang, Z. \& Zhang, S.-C.
\newblock Quantum anomalous hall effect in
  ${\mathrm{Hg}}_{1-y}{\mathrm{Mn}}_{y}\mathrm{Te}$ quantum wells.
\newblock {\em Phys. Rev. Lett.} 101, 146802 (2008).}

\bibitem{dehghani2014dissipative}{
Dehghani, H., Oka, T. \& Mitra, A.
\newblock Dissipative Floquet topological systems.
\newblock {\em Phys. Rev. B} 90, 195429 (2014).}

\bibitem{iadecola2015occupation}{
Iadecola, T., Neupert, T. \& Chamon, C.
\newblock Occupation of topological Floquet bands in open systems.
\newblock {\em Phys. Rev. B} 91, 235133 (2015).}

\bibitem{seetharam2015population}{
Seetharam, K.I., Bardyn, C.-E., Lindner, N.-H., Rudner, M.-S. \& Refael, G.
\newblock Controlled Population of Floquet-Bloch States via Coupling to Bose and Fermi Baths.
\newblock {\em Phys. Rev. X} 5, 041050 (2015).}

\bibitem{budich2015topology}{
Budich, J.C. \& Diehl, S.
\newblock Topology of density matrices.
\newblock {\em Phys. Rev. B} 91, 165140 (2015).}

\bibitem{kundu2013sumrule}
{Kundu, A. \& Seradjeh, B.
\newblock Transport Signatures of Floquet Majorana Fermions in Driven Topological Superconductors.
\newblock {\em Phys. Rev. Lett.} 111, 136402 (2013).}

\bibitem{farrell2015photon}{
Farrell, A. \& Pereg-Barnea, T.
\newblock Photon-inhibited topological transport in quantum well
  heterostructures.
\newblock {\em Phys. Rev. Lett.} 115, 106403 (2015).}

\bibitem{sobota2011unoccupied}{
Sobota, J.A., {\em et al.}
\newblock Direct Optical Coupling to an Unoccupied Dirac Surface State in the Topological Insulator Bi$_2$Se$_3$.
\newblock {\em Phys. Rev. Lett.} 111, 136802 (2013).}


\bibitem{blahaWIEN2k}{Blaha, P., Schwarz, K., Madsen, G.K.H., Kvasnicka, D. \& Luitz, J. \textit{Wien2K: An augmented plane wave and local orbitals program for calculating crystal properties}, Technische Universit{\"a}t Wien, Austria (2001).}

\bibitem{kunes2010}{Kune{\v{s}}, J.{\em et al.} Wien2wannier: From linearized augmented plane waves to maximally localized Wannier functions. {\em Comp. Phys. Comm.} 181, 1888 (2010).}

\bibitem{Mostofi2008}{Mostofi, A.A. {\em et al.} wannier90: A Tool for obtaining maximally-localised Wannier functions. {\em Comp. Phys. Comm.} 178, 685 (2008).}

\bibitem{Kormanyos2013}{Korm{\'a}nyos, A. {\em et al.}
\newblock Monolayer MoS$_2$: Trigonal warping, the $\Gamma$ valley, and
  spin-orbit coupling effects.
\newblock {\em Phys. Rev. B} 88, 045416 (2013).
}

\bibitem{Gibertini2014}{Gibertini, M., Pellegrino, F.M.D., Marzari, N. \& Polini, M.
\newblock Spin-resolved optical conductivity of two-dimensional group-VIB
  transition-metal dichalcogenides.
\newblock {\em Phys. Rev. B} 90, 245411 (2014).
}

\bibitem{Chernikov2014}{Chernikov, A., Berkelbach, T.C., Hill, H.M., Rigosi, A., Li, Y., Aslan, O.B., Reichman, D.R., Hybertsen, M.S. \& Heinz, T.F.
\newblock Exciton binding energy and nonhydrogenic Rydberg series in monolayer WS$_2$.
\newblock {\em Phys. Rev. Lett.} 113, 076802 (2014).}

\bibitem{Zhu2015}{
Zhu, B., Chen, X. \& Cui, X.
\newblock Exciton Binding Energy of Monolayer WS$_2$.
\newblock {\em Sci. Rep.} 5, 9218 (2015).}


\end{thebibliography}

\begin{thebibliography}{99}

\bibitem{Sbromley1972band}{Bromley, R.A., Murray, R.B. \& Yoffe, A.D.
The band structures of some transition metal dichalcogenides. III. group via: Trigonal prism materials.
\newblock {\em J. of Physics C: Solid State Phys.} 5, 759 (1972).}

\bibitem{Smattheis1973band}{
Mattheiss, L.F.
\newblock Band structures of transition-metal-dichalcogenide layer compounds.
\newblock {\em Phys. Rev. B} 8, 3719 (1973).}

\bibitem{Sxiao2012coupled}{
Xiao, D., Liu, G.-B., Feng, W., Xu, X. \& Yao, W.
\newblock Coupled spin and valley physics in monolayers of MoS$_2$ and other
  group-VI dichalcogenides.
\newblock {\em Phys. Rev. Lett.} 108, 196802 (2012).}

\bibitem{Szahid2013generic}{
Zahid, F., Liu, L., Zhu, Y., Wang, J. \& Guo, H.
\newblock A generic tight-binding model for monolayer, bilayer and bulk MoS$_2$.
\newblock {\em AIP Advances} 3, 052111 (2013).}

\bibitem{Scappelluti2013tight}{Cappelluti, E., Rold{\'a}n, R., Silva-Guill{\'e}n, J.A., Ordej{\'o}n, P. \& Guinea, F.
\newblock Tight-binding model and direct-gap/indirect-gap transition in
  single-layer and multilayer MoS$_2$.
\newblock {\em Phys. Rev. B} 88, 075409 (2013).}

\bibitem{Srostami2013effective}{
Rostami, H., Moghaddam, A.G. \& Asgari, R.
\newblock Effective lattice hamiltonian for monolayer MoS$_2$: Tailoring
  electronic structure with perpendicular electric and magnetic fields.
\newblock {\em Phys. Rev. B} 88, 085440 (2013).}

\bibitem{Sliu2013three}{
Liu, G.-B., Shan, W.-Y., Yao, Y., Yao, W. \& Xiao, D.
\newblock Three-band tight-binding model for monolayers of group-VIB transition
  metal dichalcogenides.
\newblock {\em Phys. Rev. B} 88, 085433 (2013).}


\bibitem{Sgibertini2014spin}{
Gibertini, M., Pellegrino, F.M.D., Marzari, N. \& Polini, M.
\newblock Spin-resolved optical conductivity of two-dimensional group-VIB
  transition-metal dichalcogenides.
\newblock {\em Phys. Rev. B} 90, 245411 (2014).}

\bibitem{Skormanyos2013monolayer}{
Korm{\'a}nyos, A. {\em et al.}
\newblock Monolayer MoS$_2$: Trigonal warping, the $\gamma$ valley, and
  spin-orbit coupling effects.
\newblock {\em Phys. Rev. B} 88, 045416 (2013).}



\bibitem{Sfeng2012intrinsic}{
Feng, W. {\em et al.}
\newblock Intrinsic spin hall effect in monolayers of group-VI dichalcogenides:
  A first-principles study.
\newblock {\em Phys. Rev. B} 86, 165108 (2012).}

\end{thebibliography}
\end{document}